\documentclass[journal]{IEEEtran}
\usepackage[T1]{fontenc}
\usepackage{cite}
\usepackage{amsmath,amssymb,amsfonts}
\usepackage{algorithm}
\usepackage{algorithmic}
\usepackage{graphicx}
\usepackage{textcomp}
\usepackage{xcolor}
\usepackage{subfigure}
\usepackage{booktabs}
\usepackage{multirow}
\usepackage{diagbox}
\usepackage{url}
\usepackage{threeparttable}
\usepackage{colortbl}
\usepackage{array}
\usepackage{textcase}
\usepackage{rotating}
\usepackage{etoolbox}

\makeatletter
\def\UrlAlphabet{%
      \do\a\do\b\do\c\do\d\do\e\do\f\do\g\do\h\do\i\do\j%
      \do\k\do\l\do\m\do\n\do\o\do\p\do\q\do\r\do\s\do\t%
      \do\u\do\v\do\w\do\x\do\y\do\z\do\A\do\B\do\C\do\D%
      \do\E\do\F\do\G\do\H\do\I\do\J\do\K\do\L\do\M\do\N%
      \do\O\do\P\do\Q\do\R\do\S\do\T\do\U\do\V\do\W\do\X%
      \do\Y\do\Z}
\def\UrlDigits{\do\1\do\2\do\3\do\4\do\5\do\6\do\7\do\8\do\9\do\0}
\g@addto@macro{\UrlBreaks}{\UrlOrds}
\g@addto@macro{\UrlBreaks}{\UrlAlphabet}
\g@addto@macro{\UrlBreaks}{\UrlDigits}

\patchcmd{\@algocf@start}
  {-1.5em}
  {0pt}
  {}{}

\makeatother

\newcommand{\SNR}[2]{${\text{SNR} #1 \text{#2}\;\text{dB}}$}
\def\post{\textit{a posteriori }}

\graphicspath{{figures/}}

\def\nt{N_{\rm t}}

\begin{document}
\setlength{\textfloatsep}{5pt}  
\setlength{\floatsep}{5pt}
\ifdefined \GramaCheck
  \newcommand{\CheckRmv}[1]{}
  \newcommand{\figref}[1]{Figure 1}%
  \newcommand{\tabref}[1]{Table 1}%
  \newcommand{\secref}[1]{Section 1}
  \newcommand{\algref}[1]{Algorithm 1}
  \renewcommand{\eqref}[1]{Equation 1}
\else
  \newcommand{\CheckRmv}[1]{#1}
  \newcommand{\figref}[1]{Fig.~\ref{#1}}%
  \newcommand{\tabref}[1]{Table~\ref{#1}}%
  \newcommand{\secref}[1]{Sec.~\ref{#1}}
  \newcommand{\algref}[1]{Algorithm~\ref{#1}}
  \renewcommand{\eqref}[1]{(\ref{#1})}
\fi
\newtheorem{theorem}{Theorem}
\newtheorem{proposition}{Proposition}
\newtheorem{assumption}{Assumption}
\newtheorem{definition}{Definition}
\newtheorem{condition}{Condition}
\newtheorem{property}{Property}
\newtheorem{remark}{Remark}
\newtheorem{lemma}{Lemma}
\newtheorem{corollary}{Corollary}
%
\title{{Reducing Pilots in Channel Estimation with Predictive Foundation Models}}

%
%
\author{Xingyu~Zhou,~\IEEEmembership{Graduate Student Member,~IEEE,}
        Le~Liang,~\IEEEmembership{Member,~IEEE,}
        Hao~Ye,~\IEEEmembership{Member,~IEEE,}
        Jing~Zhang,~\IEEEmembership{Member,~IEEE,}
        Chao-Kai~Wen,~\IEEEmembership{Fellow,~IEEE,}
        and~Shi~Jin,~\IEEEmembership{Fellow,~IEEE}
\thanks{X.~Zhou, L. Liang, J.~Zhang, and S.~Jin are with the School of Information Science and Engineering, Southeast University, Nanjing 210096, China (e-mail: \protect \url{xy_zhou@seu.edu.cn}; lliang@seu.edu.cn; jingzhang@seu.edu.cn; jinshi@seu.edu.cn).}
\thanks{H. Ye is with the Department of Electrical and Computer Engineering, University of California, Santa Cruz, CA 95064, USA
(e-mail: hye30@ucsc.edu).}
\thanks{C.-K. Wen is with the Institute of Communications Engineering, National Sun Yat-sen University, Kaohsiung 80424, Taiwan 
(e-mail: chaokai.wen@mail.nsysu.edu.tw).}
}

%
%

\maketitle

\begin{abstract}
Accurate channel state information (CSI) acquisition is essential for modern wireless systems, which becomes increasingly difficult under large antenna arrays, strict pilot overhead constraints, and diverse deployment environments. Existing artificial intelligence-based solutions often lack robustness and fail to generalize across scenarios. To address this limitation, this paper introduces a predictive-foundation-model-based channel estimation framework that enables accurate, low-overhead, and generalizable CSI acquisition. The proposed framework employs a predictive foundation model trained on large-scale cross-domain data to extract universal channel representations and provide predictive priors with strong cross-scenario transferability. A pilot processing network based on a vision transformer architecture is further designed to capture spatial, temporal, and frequency correlations from pilot observations. An efficient fusion mechanism integrates predictive priors with real-time measurements, enabling reliable CSI reconstruction even under sparse or noisy conditions. Extensive evaluations across diverse configurations demonstrate that the proposed estimator significantly outperforms both classical and data-driven baselines in accuracy, robustness, and generalization capability.

\end{abstract}

%
\IEEEpeerreviewmaketitle

\vspace{-0.4cm}
\section{Introduction}  


\IEEEPARstart{A}{ccurate} channel state information (CSI) plays a pivotal role in the design of wireless communication systems. The effectiveness of core technologies, including massive multiple-input multiple-output (MIMO) and orthogonal frequency division multiplexing (OFDM), critically depends on accurate CSI acquisition \cite{andrews20246g}. To meet the growing demand for ubiquitous connectivity and intelligent communications, wireless systems have continually scaled their antenna array dimensions, a trend expected to accelerate in sixth-generation (6G) networks \cite{wang2024tutorial}. In current fifth-generation (5G) systems, the number of antenna ports typically ranges from 16 to 128, whereas 6G is projected to scale this number to 512 and beyond \cite{shafi2025industrial,3gpp6gws2025}. 
This rapid increase in system dimensionality imposes a substantial burden on channel acquisition because pilot signals must be transmitted over increasingly dense time-frequency grids. Therefore, the development of channel estimation schemes that reduce pilot overhead while preserving high estimation accuracy has become a central research problem.

Traditional channel estimation approaches face significant limitations in adapting to dynamic and high-dimensional modern wireless systems. Linear minimum mean squared error (LMMSE)-based estimators are widely adopted due to their ability to provide the maximum \post estimate for jointly Gaussian observation models. Nonetheless, the effectiveness of LMMSE relies heavily on the accurate knowledge of channel and noise statistics, which {must be acquired} in advance and frequently adjusted to accommodate dynamic channel conditions. 
Moreover, the dependency on channel-related parameters, such as delay and Doppler spread, poses additional challenges under dynamic and heterogeneous propagation environments \cite{liu2024pd,pratik2025requestnet}. 
In addition, performing LMMSE estimation jointly over all resource elements (REs) within a resource grid entails the inversion of extremely large covariance matrices, which is computationally prohibitive.

The integration of artificial intelligence (AI) and wireless communications has {represented a transformative shift}, fundamentally reshaping the design philosophy of modern communication systems \cite{qin2024ai}. By harnessing the powerful learning and inference capabilities of neural networks (NNs), AI enables data-driven modeling, adaptation, and prediction beyond the reach of traditional analytical approaches. The shift toward AI-native wireless network design has been increasingly recognized in 3rd generation partnership project (3GPP) standardization efforts, with a particular focus on AI-enhanced physical (PHY) layer and air-interface designs in Releases 18 and beyond \cite{hoydis6GAINativeAir2021,3gpp38843study2023}. In this context, AI-based methods utilize massive wireless data, such as radio signals and CSI, to train NNs that can serve as effective alternatives to specific PHY layer modules, including beam management \cite{li2023machine}, channel prediction and estimation \cite{jiang2022accurate,zhou2025generative}, CSI feedback \cite{guo2024ai}, and localization \cite{cha20255g}. 
Furthermore, the ongoing discussions in 3GPP Release 20 highlight the urgent need for AI-enabled receiver designs that can reduce reference signal (RS) overhead \cite{R1-2505970}, aiming at building channel estimators that operate effectively across varying system configurations and numerologies.

{A variety of NN designs for CSI estimation have been explored in the literature.} For example, convolutional NNs (CNNs) were employed in \cite{soltani2019deep,li2019deep} for super-resolution and denoising by interpreting the channel response as a 2D image, thereby exploiting local temporal and frequency correlations. To track time-varying channels, recurrent neural networks (RNNs) were further introduced, leveraging their ability to model sequential dependencies in CSI evolution \cite{liao2019deep}. More recently, attention-based models and vision transformers (ViTs) \cite{vaswani2017attention,dosovitskiy2020image} were adopted to capture long-range dependencies across antenna, time, and frequency domains, thereby facilitating CSI acquisition \cite{luan2022attention,liu2024pd}.  

{Despite their promising potential, existing AI-based CSI estimation schemes face several fundamental challenges in practical air-interface deployment \cite{guo2025lvm4csi}. Most approaches rely on specialized NN architectures for performance improvement, often at the cost of scalability and deployment efficiency. Although such designs achieve strong performance in specific scenarios, their dependence on scenario-specific optimization limits generalization to unseen environments.} In practice, these models often experience significant performance degradation under distributional shifts between training and deployment conditions, making frequent retraining unavoidable. This contradicts the vision of AI-native communication systems in next-generation networks, where adaptability and autonomy are expected to be intrinsic capabilities. These limitations underscore the need for a more generalizable and unified paradigm for AI-based CSI acquisition.

Recently, large AI models (LAMs), notably large language models (LLMs) such as GPT 3 \cite{brown2020language} and DeepSeek-R1 \cite{guo2025deepseek}, have unlocked unprecedented capabilities in generalization, reasoning, and perception, motivating their exploration for PHY-layer tasks and opening up opportunities for solving the aforementioned challenges \cite{liang2025large}. 
Two complementary research directions have emerged.
{The first involves repurposing} pre-trained LLMs to communication tasks such as channel prediction \cite{liu2024llm4cp}, {beam prediction \cite{sheng2025beam},} and PHY layer multi-module joint inference \cite{zheng2024large}. These works leveraged the powerful representation and zero-shot learning capabilities of pre-trained LLMs to handle diverse scenarios without extensive task-specific redesign, hardly attainable by smaller, specific AI models.  
{The second direction focuses on constructing} wireless-specific foundation models, {trained from scratch on massive wireless datasets to provide transferable intelligence across a broad spectrum of tasks} \cite{alikhani2024large,liu2025wifo,yang2025wirelessgpt}. {Unlike general-purpose LLMs, these domain-specific models} are typically more compact, thereby enabling faster inference and adaptation. 

A salient strength of LAM-based methods lies in their proficiency in sequence modeling and next-token prediction \cite{brown2020language}, which can be leveraged to enhance the performance of the wireless inference tasks. Specifically, predictive foundation models (PFMs) \cite{nie2022time, das2024decoder} have demonstrated strong capabilities in forecasting future states in general time-series data, such as traffic, weather, and finance. These architectures have also been applied to multivariate prediction in wireless networks \cite{sheng2025wireless}, providing a unified and parameter-efficient framework for time-series forecasting across diverse communication tasks. 
{Given the strong temporal correlations inherent in fading channels, these models offer a unique opportunity to exploit historical CSI as a source of side information. By treating the forecasted channel states as \textit{predictive priors}, it becomes feasible to augment the instantaneous information captured by sparse pilots.}
This motivates an important research question: \textit{Can predictive priors derived from foundation models and historical CSI be synergistically combined with pilot information, thereby effectively reducing pilot overhead while maintaining high estimation accuracy?}

In response, this work adapts PFMs to the characteristics of wireless channels and proposes a PFM-aided channel estimation framework. {Distinct from conventional methods that rely solely on current observations, our approach employs a ``predict-and-refine'' strategy.
The PFM first exploits historical CSI to generate forecasted channel states as predictive priors, serving as a highly informative baseline. The instantaneous pilot information is then utilized for calibrating the forecast, rather than estimating from scratch.} This methodology establishes a foundational channel estimation model that generalizes across a wide range of configurations and scenarios. The contributions of this paper are summarized as follows.

\begin{itemize}
  \item \textbf{Learning a Channel Estimation-Specific Predictive Foundation Model:} We construct a specialized PFM tailored for wireless channel estimation, leveraging the pre-trained time-series forecasting model and finetuning it with domain-specific adaptation. The pre-trained weights show effectiveness in channel acquisition, and the finetuned model effectively transfers cross-domain predictive knowledge, significantly enhancing generalization across propagation environments and system configurations. 
  
  \item \textbf{Channel Estimation Framework and Workflow With a Predictive Foundation Model:} We develop a PFM-aided channel estimation framework that establishes a systematic workflow for integrating predictive priors from the PFM with pilot-based estimation using an efficient fusion mechanism. By leveraging these predictive priors inferred from historical CSI to complement scarce online measurements, this framework enables accurate, low-overhead, and generalizable channel estimation. 
  
  \item \textbf{Advanced NN Design for {Pilot Information} Processing:}  
  An advanced neural architecture is designed to capture temporal, frequency, and spatial CSI correlations based on pilot observations. By incorporating the ViT architecture and attention modules, the model learns rich multi-scale representations that improve channel reconstruction fidelity under sparse or noisy pilot conditions.

  \item \textbf{Comprehensive Numerical Validation:} Extensive experiments across diverse channel configurations demonstrate that the proposed PFM-aided estimator consistently outperforms conventional and data-driven baselines in both estimation accuracy and generalization, particularly under sparse pilot patterns, paving the way toward foundation model-based approaches for efficient and intelligent wireless channel acquisition.

\end{itemize}

\textit{Notations:} 
For any matrix $\mathbf{A}$, $\mathbf{A}^{\top}$ and $\mathbf{A}^{\rm H}$ denote the transpose and conjugate transpose of $\mathbf{A}$, respectively. 
Also, 
$\|\cdot\|_2$ is the $l_2$-norm, 
$\|\cdot\|_F$ is the Frobenius norm, 
$\emptyset$ is the empty set, 
$\odot$ is the Hadamard product, and
$\mathbb{E}[\cdot]$ is the expectation operator.  
Moreover, $\mathbb{Z}^+$, $\mathbb{R}$, and $\mathbb{C}$ denote the sets of positive integers, real numbers, and complex numbers, respectively.


\vspace{-0.1cm}
\section{Problem Formulation and Preliminaries} \label{sec:problem}  

\subsection{Channel Estimation Problem} 
We consider the downlink transmission between a BS with $\nt$ antennas and a single-antenna user equipment (UE) utilizing OFDM modulation.{\footnote{Although we assume downlink transmission with a single-antenna UE in the system model for simplicity, the proposed approach can be seamlessly applied to uplink transmission and systems with multi-antenna UEs.}}
The system operates under the 5G new radio (NR) numerology, where each frame consists of 10 subframes, and each subframe is further partitioned into several slots depending on the subcarrier spacing. Each slot {$i\in\mathbb{Z}^+$} comprises $T$ OFDM symbols in the time domain and $K$ subcarriers in the frequency domain, forming a time-frequency resource grid. A group of 12 consecutive subcarriers constitutes a resource block (RB), and each time-frequency unit within the grid is referred to as {an} RE. 

Channel estimation relies on RS, which comprises predefined pilot symbols occupying dedicated REs within each RB. Let $K_{\rm P}$ and $T_{\rm P}$ denote the number of subcarriers and symbols allocated for pilot transmission within a slot. To simplify analysis, we assume wideband and identity precoding at the transmitter during pilot transmission, i.e., an identity matrix is adopted as the precoder across all RBs. The UE's received signal over the pilot RE positions can be written as
\CheckRmv{
  \begin{equation}
    \mathbf{Y}_{\rm P} = \sum_{n=1}^{\nt} \mathbf{H}_{\rm P}^{[n]} \odot \mathbf{X}_{\rm P}^{[n]} + \mathbf{N}_{\rm P},
  \end{equation}
} 
where $\mathbf{H}_{\rm P}^{[n]}\in \mathbb{C}^{K_{\rm P} \times T_{\rm P}}$ and ${\mathbf{X}_{\rm P}^{[n]}\in \mathbb{C}^{K_{\rm P} \times T_{\rm P}}}$ denote the channel frequency response and transmitted pilot symbols from the $n$-th transmit antenna at the pilot positions, respectively, and $\mathbf{N}_{\rm P}$ represents the additive white Gaussian noise (AWGN) with variance $\sigma^2$. Following the 5G NR specifications, different transmit {antennas} employ orthogonal pilot patterns by means of frequency-division multiplexing (FDM) or orthogonal cover codes, ensuring separability of {per-antenna} channels at the receiver. 

{For each transmission slot $i$, the goal of channel estimation is to obtain the CSI across all REs and antennas, denoted as $\mathbf{H}^{(i)} \in \mathbb{C}^{\nt K \times T}$, given the known pilot symbols from all transmit antennas, $\mathbf{P}^{(i)} \in \mathbb{C}^{\nt \times K_{\rm P} \times T_{\rm P}}$, and the corresponding noisy pilot observations, $\mathbf{Y}_{\rm P}^{(i)}\in \mathbb{C}^{K_{\rm P} \times T_{\rm P}}$. Here, we have
\CheckRmv{
  \begin{equation}
    \mathbf{H}^{(i)} \triangleq [\mathbf{h}_1^{(i)},\mathbf{h}_2^{(i)},\ldots,\mathbf{h}_T^{(i)}],
  \end{equation}
}
where $\mathbf{h}_t^{(i)} \in \mathbb{C}^{\nt K\times 1}$, $ t\in \{1,\ldots, T\}$, denotes the vectorized spatial-frequency CSI at the $t$-th symbol of slot $i$.
As pilot transmission incurs non-negligible overhead, a key challenge lies in acquiring accurate CSI under low pilot density. To address this issue, our proposed approach integrates sparse pilot observations with predictive priors from a PFM, enabling accurate and robust CSI reconstruction across diverse configurations and channel conditions. To quantify the accuracy of the estimated CSI, the normalized mean squared error (NMSE) is adopted as the performance metric, defined as
\CheckRmv{
  \begin{equation}
    \text{NMSE}=\mathbb{E}\Bigg[\frac{\big\|\hat{\mathbf{H}}^{(i)} - {\mathbf{H}}^{(i)}\big\|_F^2}{\big\|{\mathbf{H}}^{(i)}\big\|_F^2}\Bigg], \label{eq:nmse}
  \end{equation}
}
where $\hat{\mathbf{H}}^{(i)}\in \mathbb{C}^{\nt K\times T}$ denotes the estimated CSI for slot $i$.}

\subsection{Time-Series Predictive Foundation Model} \label{sec:pfm_basic}  
Inspired by the advances of LLMs in natural language processing (NLP), PFMs for time-series forecasting have received increasing research attention \cite{nie2022time,das2024decoder}. Let $\mathbf{s}_t \in \mathbb{R}^{M \times 1}$ denote the observation vector of a multivariate time series at time step $t$. Such models aim to map a historical context of length $L$ to a future horizon of length $S$, that is, to predict the next $S$ time steps based on the past $L$ steps:
\CheckRmv{
  \begin{equation}
    f_{\boldsymbol{\theta}}:(\mathbf{S}) \longrightarrow \hat{\mathbf{S}},
  \end{equation}
} 
where $f_{\boldsymbol{\theta}}$ represents the PFM parameterized by $\boldsymbol{\theta}$, $\mathbf{S} = {[\mathbf{s}_{t_0-L+1},\ldots,\mathbf{s}_{t_0}] \in \mathbb{R}^{M\times L}}$ denotes the historical context with $t_0$ being the current time step index, and $\hat{\mathbf{S}} = {[\hat{\mathbf{s}}_{t_0+1},\ldots,\hat{\mathbf{s}}_{t_0+S}] \in \mathbb{R}^{M\times S}}$ represents the predicted future values.

Motivated by the forecasting and generalization capabilities of PFMs, we consider their application to the task of wireless channel acquisition.
To align the PFM capability with the slot-based transmission structure, we configure the model input length and prediction horizon to match the slot duration, i.e., $L=T$ and $S=T$. Consequently, the PFM serves as a slot-predictor that maps the spatial-frequency CSI of the previous slot, denoted as $\mathbf{H}^{(i-1)} \in \mathbb{C}^{\nt K \times T}$, to that of the current slot, $\mathbf{H}^{(i)} \in \mathbb{C}^{\nt K \times T}$. Hence, the channel prediction problem based purely on historical CSI can be formulated as 
\CheckRmv{
  \begin{equation}
    {f_{\boldsymbol{\theta}}:(\mathbf{H}^{(i-1)}) \longrightarrow \hat{\mathbf{H}}^{(i)}.} \label{eq:p1}
  \end{equation}
}

\section{Proposed Methods} \label{sec:method}  
{In this section, we first present} the overall framework and the corresponding workflow of the proposed channel estimator. {We then} elaborate on the architectures of the PFM and the pilot processing ViT within the proposed framework. Finally, we introduce the fusion module and detail the training procedure of the integrated model. 

\subsection{Overall Framework and Workflow}  \label{sec:overall}  

\CheckRmv{
  \begin{figure*}[t]
    \centering
      \subfigure[Overall framework]{
        \includegraphics[width=4in]{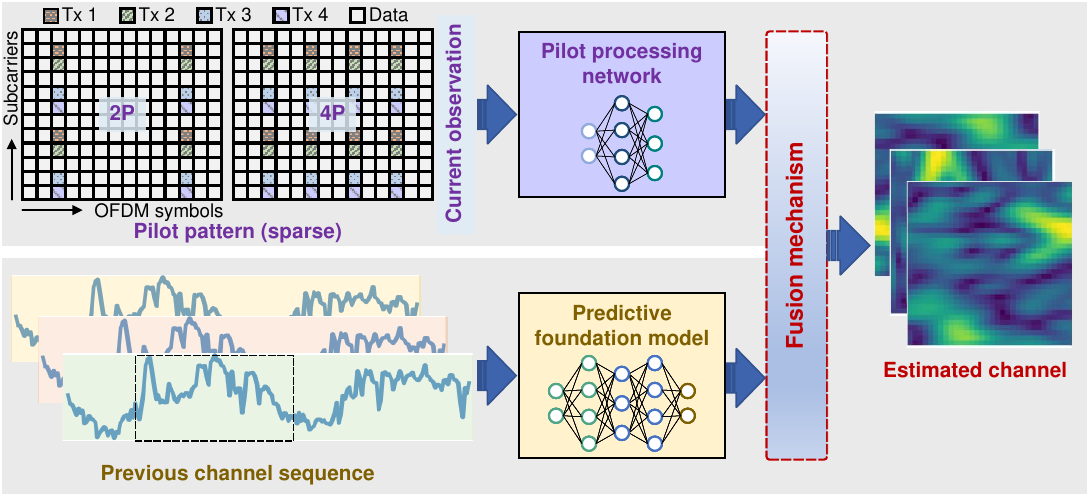}
        \label{fig:overall}
      }
      \subfigure[{Workflow}]{
        \includegraphics[width=2.9in]{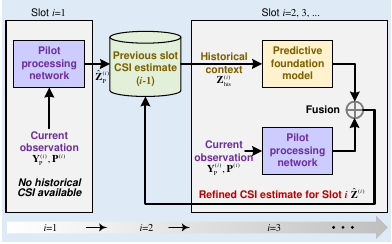}
        \label{fig:workflow}
      }
    \caption{Overall framework and {slot-based} workflow of the proposed PFM-aided channel estimator.}
    \label{fig:framework}
  \end{figure*}
}

The overall framework and workflow of the proposed channel estimator are presented in \figref{fig:framework}. As depicted in \figref{fig:overall}, the framework includes two {major branches: (i) predictive foundation models and (ii) pilot information processing.} Each branch generates a complementary representation of the current CSI, and a dedicated fusion mechanism is designed to merge these representations into the final channel estimate.

In the first branch, {the PFM $f_{\boldsymbol{\theta}}$ described in Section~\ref{sec:pfm_basic}} is constructed to leverage \textit{historical CSI} and generate auxiliary predictive information on the current channel state. Unlike traditional estimators that rely exclusively on pilot observations, this PFM introduces additional information that enhances CSI acquisition. Furthermore, owing to the strong generalization capabilities of PFM, the proposed estimator exhibits robust performance even when deployed in previously unseen scenarios. 
In the second branch, a pilot information processing NN, denoted as $g_{\boldsymbol{\Omega}}$ (parameterized by $\boldsymbol{\Omega}$), is developed to obtain a CSI estimate based on the received pilot observations. This component plays a role analogous to that of conventional pilot-based estimators, providing reliable \textit{instantaneous information determined solely by the current slot}.
To effectively combine the strengths of these two branches, we design a fusion mechanism that operates on the \textit{hidden representations} extracted from both the PFM and the pilot processing network. 
Detailed designs are presented in Sections \ref{sec:pfm} {to} \ref{sec:fusion}.

{Applying the proposed framework to the channel estimation problem requires resolving several practical issues. 
Specifically, the formulation in \eqref{eq:p1} presumes that the PFM has access to perfect historical CSI, which is unattainable in practice and thus requires an alternative construction of the historical context.} To this end, we develop a slot-based workflow as depicted in \figref{fig:workflow}. The key design principle is a recursive estimation process where the \textit{output} of slot $(i-1)$ serves as the \textit{historical input} for slot $i$. 
To formalize this recursive workflow and facilitate NN processing, let $\mathbf{Z}^{(i)}\in\mathbb{R}^{2\nt K \times T}$ and $\hat{\mathbf{Z}}^{(i)}\in\mathbb{R}^{2\nt K \times T}$ denote the real-valued tensor representations of $\mathbf{H}^{(i)}$ and $\hat{\mathbf{H}}^{(i)}$, respectively, {which are obtained by concatenating their real and imaginary components along the antenna dimension.} Furthermore, let $\mathbf{Z}_{\rm his}^{(i)}$ denote the historical input fed into the PFM at slot $i$.
{Considering the transmission over multiple slots, as shown in \figref{fig:workflow}, we have the following two cases}.
\begin{enumerate}
  \item \textbf{Initial Slot ($i=1$):} No historical CSI is available, and $\mathbf{Z}_{\rm his}^{(i)}$ becomes $\emptyset$. The estimation relies solely on the pilot processing network. The estimated CSI is given by
  \CheckRmv{
    \begin{equation}
      \hat{\mathbf{Z}}^{(i)} = \hat{\mathbf{Z}}_{\rm P}^{(i)} = g_{\boldsymbol{\Omega}}(\mathbf{Y}_{\rm P}^{(i)}, \mathbf{P}^{(i)}),
    \end{equation}
  }
  where $\hat{\mathbf{Z}}_{\rm P}^{(i)}\in\mathbb{R}^{2\nt K \times T}$ is the output of $g_{\boldsymbol{\Omega}}$ at slot $i$. 
  \item \textbf{Subsequent Slots ($i\geq 2$):} The estimated CSI from the previous slot, $\hat{\mathbf{Z}}^{(i-1)}$, is utilized as the historical context, $\mathbf{Z}_{\rm his}^{(i)}$, for the PFM. {By incorporating both the historical CSI and the instantaneous pilot information, the final CSI estimate for the current slot is given by
  \CheckRmv{
    \begin{equation}
      \hat{\mathbf{Z}}^{(i)} = F_{\boldsymbol{\Theta}}(\mathbf{Z}_{\rm his}^{(i)}, \mathbf{Y}_{\rm P}^{(i)}, \mathbf{P}^{(i)}),
    \end{equation}
  }
  where $F_{\boldsymbol{\Theta}}$ denotes the proposed joint estimator that integrates the two branches and is parameterized by $\boldsymbol{\Theta}$. This formulation effectively refines the PFM-based prediction using current pilot observations.}
\end{enumerate} 
Based on this workflow, the historical input used at each slot $i$ can be summarized as 
\CheckRmv{
  \begin{equation}
    \mathbf{Z}_{{\rm his}}^{(i)}=\left\{\begin{array}{ll}
      \emptyset, & \text { if } i=1, \\
      {\hat{\mathbf{Z}}^{(i-1)}}, & \text { if } i\geq 2. 
    \end{array}\right.
    \label{eq:his}
  \end{equation}
}   
\begin{remark}
  One might argue that extending the historical context to include multiple prior slots (e.g., $i-2, i-3$) could provide richer temporal information. However, the coherence time of practical wireless channels is limited. Empirical observations suggest that CSI from distant slots becomes {less correlated}. Incorporating such outdated information not only increases computational overhead but also exacerbates error propagation, as artifacts from earlier estimations accumulate, ultimately degrading estimation performance. Therefore, we restrict the history to the immediately preceding slot ($i-1$). 
\end{remark}

\subsection{{Predictive Foundation Model for Channel Estimation}}  \label{sec:pfm}  

\CheckRmv{
  \begin{figure*}[t]
    \centering
    \includegraphics[width=6.2in]{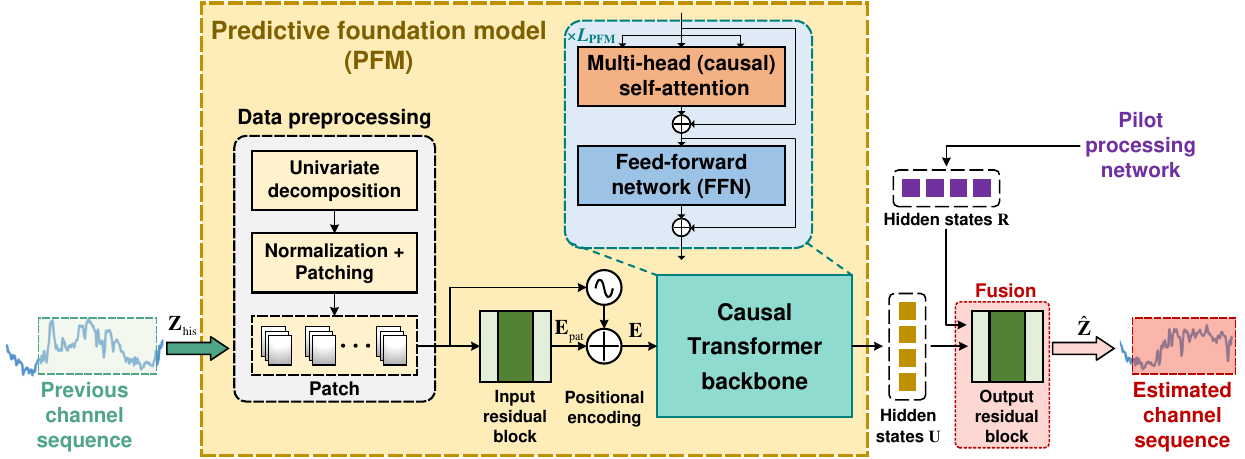}
    \caption{Architecture of the PFM tailored to channel estimation. The model consists of data preprocessing, an input residual block, positional encoding, and a causal transformer backbone, followed by an output residual block. 
    }
    \label{fig:pfm}
  \end{figure*}
}

The architecture of the PFM \cite{das2024decoder} is presented in \figref{fig:pfm}. {Its primary operations include} data preprocessing, input projection, positional encoding, causal transformer-based feature extraction, and output projection.  These components collectively form the predictive pipeline, as detailed below. {For notational simplicity, the slot index $(\cdot)^{(i)}$ is omitted in the illustration.}

\subsubsection{\textbf{Data Preprocessing}}

The data preprocessing pipeline contains the following steps: 
{\begin{itemize}
  \item \textit{Univariate decomposition.} It is demonstrated in \cite{nie2022time} that performing prediction for each variable separately within a multivariate time-series surpasses approaches that jointly predict multiple variables. 
  To align with this established practice, univariate decomposition is first conducted, decomposing the input $\mathbf{Z}_{\rm his}$ into $Q\triangleq2\nt K$ univariate historical sequences, 
  \CheckRmv{
    \begin{equation}
      [\mathbf{z}^{[1]}, \ldots, \mathbf{z}^{[Q]}]^{\top} = \operatorname{RowSplit}(\mathbf{Z}_{\rm his}),
    \end{equation}
  }
  where $\mathbf{z}^{[q]}\in \mathbb{R}^{T\times 1}$ ($q\in\{1,\ldots, Q\}$) denotes the historical sequence extracted from the $q$-th channel of the multivariate time-series. Each of these univariate sequences is then fed into the shared transformer and processed through distinct forward passes. 
  \item \textit{Normalization and patching.} 
  To mitigate distributional shifts {across channel realizations} and enhance model robustness, each univariate input sequence $\mathbf{z}^{[q]}$ is normalized by subtracting its sample mean and scaling by its standard deviation, yielding the normalized sequence. 
  After normalization, the patching technique is applied similarly as in NLP tasks, which breaks down the normalized sequence into non-overlapping patches of length $L_{\rm pat}$, forming the input matrix $\mathbf{Z}_{\rm pat}^{[q]} \in \mathbb{R}^{N_{\rm pat} \times L_{\rm pat}}$, where $N_{\rm pat} = \left\lfloor T / L_{\rm pat} \right\rfloor $ is the number of patches. 
  The generated patches, analogous to tokens in NLP, capture localized semantic features within the time-series data. This patching mechanism also reduces the token count by a factor of $L_{\rm pat}$, thereby achieving notable savings in both computational and memory complexity.
  
\end{itemize}}

\subsubsection{\textbf{Input Projection and Positional Encoding}}
After preprocessing, the generated patches are projected to a high-dimensional latent space corresponding to the transformer backbone via a residual block \cite{he2016deep}:  
\CheckRmv{
  \begin{equation}
    \mathbf{E}_{\rm pat} = \operatorname{InputResBlock}(\mathbf{Z}_{\rm pat}), \label{eq:input_proj}
  \end{equation}
}
where $\mathbf{E}_{\rm pat} \in \mathbb{R}^{N_{\rm pat} \times d}$ denotes the patch embeddings, and $d$ is the latent space dimension. {Note that the superscript $(\cdot)^{[q]}$ is omitted since each univariate sequence is independently processed.}
To preserve the temporal order information of the patches, positional encodings are incorporated into the patch embeddings prior to their input into the transformer\cite{vaswani2017attention}:
\CheckRmv{
  \begin{equation}
   \mathbf{E} = \mathbf{E}_{\rm pat} + \operatorname{PE}(\mathbf{m}), 
  \end{equation}
}
where $\operatorname{PE}(\cdot)$ denotes the deterministic sinusoidal positional encoding function, and $\mathbf{m}\in \mathbb{R}^{N_{\rm pat} \times 1}$ {is} the patch index vector. {The resulting tensor $\mathbf{E}\in \mathbb{R}^{N_{\rm pat} \times d}$ serves as the input to the transformer}.

{\subsubsection{\textbf{Causal Transformer Backbone}} \label{sec:transformer} 
The core of the PFM is a standard causal transformer backbone employed to extract high-level representations from the input. {The backbone} contains a stack of $L_{\rm PFM}$ transformer layers, {each containing} a multi-head \textit{causal} self-attention module followed by a feed-forward network (FFN), as presented in \figref{fig:pfm}. We utilize $N_{\rm H}$ attention heads with a causal mask to ensure that each token attends only to its current and past tokens, preventing information leakage from future tokens. The mathematical formulation of the causal attention mechanism is detailed in Appendix~\ref{app:causal}. The aggregated output of the attention heads is processed by the FFN, yielding the layer output.  
After propagating through all the stacked transformer layers, high-level feature representations, $\mathbf{U} = [\mathbf{u}_{1}, \ldots, \mathbf{u}_{N_{\rm pat}}]^{\top} \in \mathbb{R}^{N_{\rm pat} \times d}$, are ultimately derived as the output of the transformer backbone.}

\subsubsection{\textbf{Output Projection and Denormalization}}
The final {step} is to project the output feature of the transformer backbone to the prediction. A decoder-only mode \cite{brown2020language} is adopted, wherein future values are predicted based on all {preceding} input patches in a given sequence. Notably, the last patch representation within $\mathbf{U}$, namely $\mathbf{u}_{N_{\rm pat}}\in \mathbb{R}^{d\times 1}$, encapsulates contextual information aggregated from all previous patches. Therefore, only $\mathbf{u}_{N_{\rm pat}}$ is utilized for output generation, thereby maintaining scalability to a varying number of patches. Specifically, an output residual block is employed to map $\mathbf{u}_{N_{\rm pat}}$ to the $T$-point normalized prediction vector $\hat{\mathbf{z}}_{\rm norm}\in \mathbb{R}^{T \times 1}$:
\CheckRmv{
  \begin{equation}
    \hat{\mathbf{z}}_{\rm norm}= \operatorname{OutputResBlock}(\mathbf{u}_{N_{\rm pat}}).
  \end{equation}
}
This direct prediction of the full horizon demonstrates enhanced accuracy and reduced iteration overhead compared to multi-step autoregressive inference for prediction \cite{zeng2023transformers}. The final prediction $\hat{\mathbf{z}}\in \mathbb{R}^{T\times 1}$ is derived by denormalizing $\hat{\mathbf{z}}_{\rm norm}$ using the mean and standard deviation applied during preprocessing. {Predictions across}  all antennas and subcarriers are {then concatenated to obtain} the complete {predicted} CSI {$\hat{\mathbf{Z}}_{\rm PFM}\in \mathbb{R}^{Q\times T}$}.

{\begin{remark}
  The cross-domain transferability of the PFM from macroscopic data, e.g., weather and traffic, to microscopic channel fading is grounded in the mathematical isomorphism of temporal dynamics. Despite vastly different physical time scales, both domains are fundamentally governed by harmonic superpositions and autoregressive correlations, allowing the pre-trained PFM to act as a universal temporal feature extractor. Moreover, the sequence-wise normalization effectively abstracts away absolute physical time units and amplitude scales, rendering the inputs scale-invariant. Consequently, the pre-trained PFM seamlessly serves as a robust structural prior for the wireless domain, ready to be refined via domain-specific fine-tuning.
\end{remark}}

\CheckRmv{
  \begin{figure*}[t]
    \centering
      \subfigure[Overall architecture]{
        \includegraphics[width=2.65in]{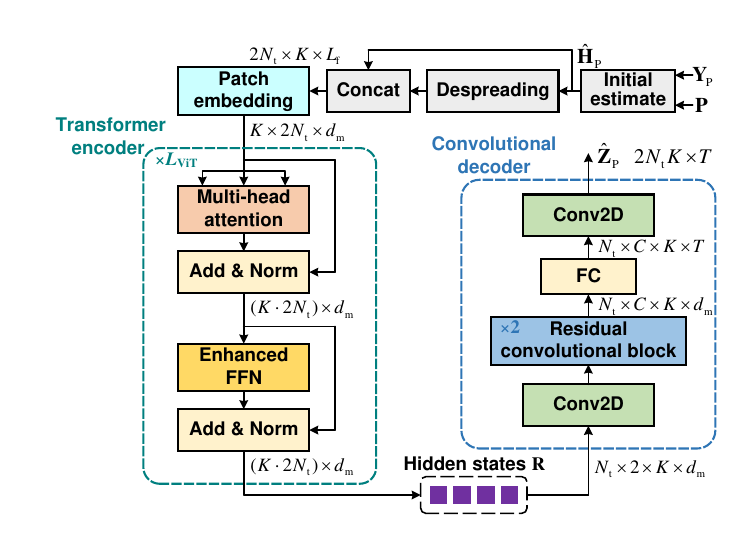}
        \label{fig:vit_overall}
      }
      \subfigure[Operation of the enhanced FFN]{
        \includegraphics[width=4in]{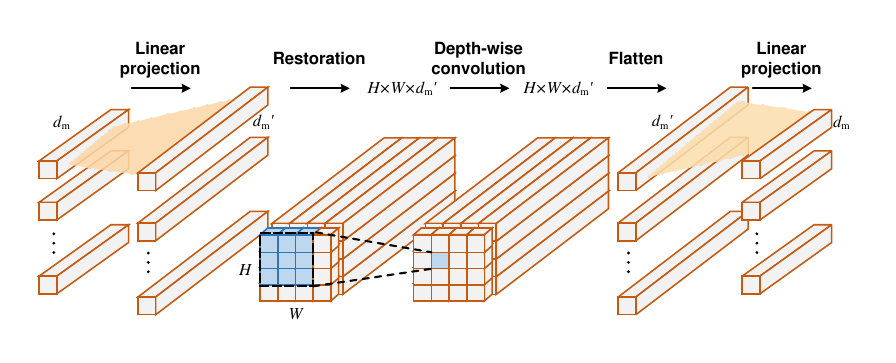}
        \label{fig:effn}
      }
    \caption{Architecture of the proposed ViT-based pilot processing network.}
    \label{fig:vit}
  \end{figure*}
}

\subsection{Pilot Information Processing Network} \label{sec:pilot} 

The structure of the proposed ViT $g_{\boldsymbol{\Omega}}$ for pilot information processing is demonstrated in \figref{fig:vit}. 
Following \cite{luan2022attention}, the overall structure {in \figref{fig:vit_overall}} contains a transformer encoder for encoding spatial, frequency dependency in the hidden representation for fusion and a residual decoder for capturing time-frequency correlations in the estimated results. {The components of this network are outlined below, and the implementation details can be found in Appendix~\ref{app:pilot}.}

\subsubsection{\textbf{Input Feature Preparation}} 
We first derive an initial coarse estimate, {$\hat{\mathbf{H}}_{\rm P} \in \mathbb{R}^{2\nt\times K \times T}$, based on the least squares (LS) method and interpolation, where the real and imaginary parts are concatenated along the antenna dimension.} 
This estimate serves as a noisy approximation of the true CSI. {By utilizing the interpolated estimate rather than raw pilot observations, the input tensor to the ViT consistently maintains a full-grid dimension of $2\nt \times K \times T$. This unified input representation decouples the network from specific pilot RE indices, thereby ensuring scalability to varying pilot patterns without retraining.} To further suppress the noise and leverage its zero-mean property, a despreading operation \cite{3gpp38211} is adopted to average the estimate across $R_{\rm T}$ and $R_{\rm F}$ neighboring REs in the time and frequency domains, respectively, thereby enhancing feature consistency. {The feature maps before and after the despreading operation are concatenated along the time dimension, resulting in a feature tensor of size $2\nt \times K \times L_{\rm f}$, where $L_{\rm f}=T+T/R_{\rm T}$. To capture spatial correlations, the features within this tensor are grouped by subcarrier. Specifically, the spatial-time domain feature on each subcarrier is partitioned into $2\nt$ patches of length $L_{\rm f}$, which are then linearly projected onto a latent space of dimension $d_{\rm m}$ via a fully-connected (FC) layer. This yields the input embeddings for the subsequent transformer encoder.}

\subsubsection{\textbf{Transformer Encoder}} 
{The transformer encoder is composed of $L_{\rm ViT}$ layers. Within each layer, the self-attention mechanism with $N_{\rm H}^{\prime}$ heads operates bi-directionally without the causal mask mentioned in Section~\ref{sec:pfm}, allowing all patches on each subcarrier to attend to each other and thereby capturing global dependencies.\footnote{Note that the patch embedding on each subcarrier is processed individually in the attention module to avoid the high complexity brought by an extremely large number of patches.} Moreover, to better capture local frequency correlations that are overlooked by the attention module, we replace the standard FFN in each layer with an enhanced FFN, as illustrated in \figref{fig:effn}.

The enhanced FFN incorporates a depth-wise convolution (DWConv) \cite{yuan2021incorporating}. The token sequence is first linearly projected to a higher dimension of $d_{\rm m}^{\prime}$ and then restored into a 2D spatial-frequency grid based on the original positions of the tokens. A $3\times 3$ DWConv is subsequently applied, as illustrated in the dashed boundary area in \figref{fig:effn}, to extract local features among neighboring tokens. The convolved features are then flattened and linearly projected back to the original dimension $d_{\rm m}$, yielding the FFN output. Detailed mathematical formulation of this enhanced FFN is presented in Appendix~\ref{app:pilot}.}

Note that each linear projection and DWConv operation is followed by an adaptive layer normalization (AdaLN) and a Gaussian error linear unit (GELU) activation. {Unlike} conventional layer normalization that uses learnable yet inference-time fixed \textit{dimension-wise} scale and shift parameters of size $H\times W$, the AdaLN operation utilizes \textit{depth-wise} scale and shift parameters of size $d_{\rm m}$ ({or $d_{\rm m}^{\prime}$}) that adaptively change with the input conditions \cite{peebles2023scalable}. This design maintains the representational flexibility of standard layer normalization via learnable parameters while ensuring scalability to varying numbers of antennas $\nt$ and subcarriers $K$, making it suitable for practical deployment under dynamic system configurations. The AWGN variance $\sigma^2$ is {used} as the input condition in our implementation to enhance adaptability to varying signal-to-noise power ratios (SNRs).

{By using the attention modules and enhanced FFNs, the stacked transformer layers finally produce the hidden states $\mathbf{R} \in \mathbb{R}^{Q \times d_{\rm m}}$ as the output of the encoder, where $Q=K \cdot 2\nt$.}

\subsubsection{\textbf{Convolutional Decoder}} 
{The decoder enhances feature fusion in the time-frequency domain and reconstructs the channel estimate from the encoder's hidden representations. It consists of an initial convolutional layer ($C=32$ filters, $5\times 5$ kernels), followed by two residual convolutional blocks. 
Each block comprises two convolutional layers ($C=32$ filters, $5\times 5$ kernels) with rectified linear unit activation.
The final output is generated by an FC layer coupled with a convolutional layer to restore the original time-domain dimension $T$, yielding the final pilot-based estimate $\hat{\mathbf{Z}}_{\rm P} \in \mathbb{R}^{Q \times T}$.}

\subsection{{Fusion Module}} \label{sec:fusion}
{As outlined in Section~\ref{sec:overall}, the PFM described in Section~\ref{sec:pfm} is further combined with the pilot information processing output through a fusion module.} 
As demonstrated in \figref{fig:pfm}, the hidden states from the two branches are utilized for fusion, and the output residual block of the PFM is repurposed as the fusion module.\footnote{This choice of information fusion via the output projection layers is both convenient and commonly adopted in practice. It is also empirically observed that this approach outperforms weighting- or attention-based fusion schemes.}

Specifically, for the PFM, the last patch hidden representations of all $Q$ univariate sequences (from all antennas and subcarriers) are concatenated to form the hidden state
\CheckRmv{
  \begin{equation}
    \mathbf{U}^{\prime}=[\mathbf{u}_{N_{\rm pat}}^{[1]}, \ldots, \mathbf{u}_{N_{\rm pat}}^{[Q]}]^{\top} \in \mathbb{R}^{Q\times d},
  \end{equation}
} 
which aggregates information from all preceding patches. This hidden state is then combined with the output state of the pilot processing ViT's encoder, $\mathbf{R}$, yielding $\mathbf{U}^{\prime\prime} = [\mathbf{U}^{\prime}, \mathbf{R}] \in \mathbb{R}^{Q\times (d+d_{\rm m})}$. 
Finally, the fusion residual block projects this joint feature to the CSI domain as   
\CheckRmv{
  \begin{equation}
    \hat{\mathbf{Z}} = \operatorname{FusionResBlock}(\mathbf{U}^{\prime\prime}) \in \mathbb{R}^{Q\times T}, \label{eq:fusion}
  \end{equation}
}
which corresponds to the estimated CSI across all antennas and subcarriers within a slot.

{To clarify the dimensional operations in \eqref{eq:fusion}, the tensor $\mathbf{U}^{\prime\prime}$ contains $Q$ independent univariate sequences. The fusion residual block treats $Q$ as a parallel batch dimension, independently transforming each sequence's feature dimension to yield the $Q\times T$ output. Furthermore, ``repurposing'' the PFM's output block indicates that the fusion module inherits its identical structural design, e.g., residual connections and activations. However, to accommodate the joint representation, a dimensional adaptation is employed: the input dimension of its first linear layer expands from $d$ to $d+d_{\rm m}$, properly mapping the concatenated hidden states to the $T$-point CSI estimates.}

\subsection{Model Training}  \label{sec:train}  

\CheckRmv{
  \begin{figure}[t]
    \centering
    \includegraphics[width=2.7in]{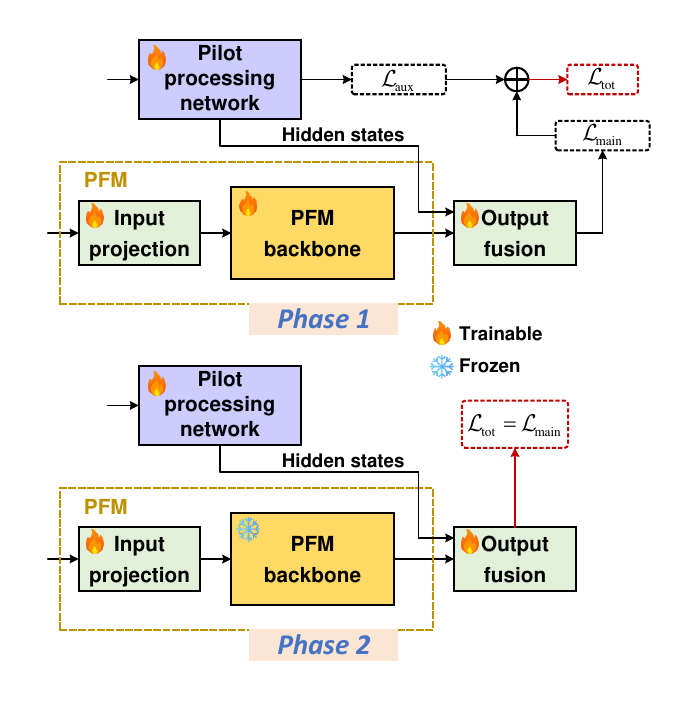}
    \caption{Illustration of the proposed two-phase training strategy.}
    \label{fig:training}
  \end{figure}
}

The model training primarily follows the workflow outlined in \figref{fig:workflow}. 
Each training sample contains a triplet of previous and current channel realizations, denoted as $\{{\hat{\mathbf{H}}_{\rm P}^{\langle l \rangle}}, \mathbf{Z}_{\rm his}^{\langle l \rangle}, \mathbf{Z}^{\langle l \rangle}\}$, where $l$ is the sample index, {${\hat{\mathbf{H}}_{\rm P}^{\langle l \rangle}\in \mathbb{R}^{2\nt\times K \times T}}$} and $\mathbf{Z}_{\rm his}^{\langle l \rangle}\in \mathbb{R}^{Q \times T}$ are the input features, and $\mathbf{Z}^{\langle l \rangle}\in \mathbb{R}^{Q \times T}$ is the label. Specifically, $\hat{\mathbf{H}}_{\rm P}^{\langle l \rangle}$ denotes the initial coarse estimate of the CSI in the current slot, {and $\mathbf{Z}_{\rm his}^{\langle l \rangle}$ denotes the estimated CSI of the previous slot}. To enhance generalization across diverse configurations, we improve the input diversity by providing  LS-based initial estimates with both linear and frequency-time LMMSE interpolation under varying pilot patterns, SNRs, and channel conditions, as detailed in Section~\ref{sec:simu_setup}. {Moreover, generating the historical trajectory} based on \eqref{eq:his} involves multistep iterative propagation, where the model's predictions are recursively fed back into itself. Such recursive generation significantly increases training complexity and often leads to instability. To mitigate these issues, we replace the recursively generated trajectory with LMMSE-based estimates, thereby supplying the model with sufficiently accurate historical CSI that matches the quality expected in the intended inference workflow.
Empirically, this substitution preserves the fidelity of historical information, facilitates stable training of the PFM, and incurs virtually no performance degradation relative to using fully workflow-matched trajectories during training.  

{Furthermore, to construct a robust estimator, we first adapt the PFM backbone to the wireless channel domain, followed by a two-phase training strategy for the complete model, as depicted in \figref{fig:training}. The Adam optimizer is adopted throughout the entire training process.

\textbf{Predictive Foundation Model Adaptation:} Prior to the end-to-end training, the PFM backbone is initialized using the pre-trained weights from \cite{das2024decoder} and then finetuned on the channel dataset. This adaptation step is crucial to align the features learned from large-scale pre-training with the properties of wireless channels, thereby constructing a foundational predictive model for the wireless channel domain. The loss function for this adaptation is defined as the mean squared error (MSE) between the PFM's prediction $\hat{\mathbf{Z}}_{\rm PFM}$ and the ground-truth channel tensor $\mathbf{Z} \in \mathbb{R}^{Q \times T} $ of the current slot:
\CheckRmv{
  \begin{equation}
    \mathcal{L}_{\rm PFM} = \operatorname{MSE}(\hat{\mathbf{Z}}_{\rm PFM}, \mathbf{Z}),
  \end{equation}
}
averaged over all samples (antennas, subcarriers, and slots) within a mini-batch. 
We employ a small learning rate $l_1$ and a large weight decay factor $\lambda_1$ to enable a stable adaptation, providing a reliable starting point for the subsequent estimator training.

\textbf{Two-Phase Estimator Training:} 
Building upon the adapted backbone, we optimize the entire estimator through two training phases. In the first phase, the entire model, including the adapted PFM and the pilot processing network, is jointly optimized using learning rate $l_1$ and weight decay factor $\lambda_1$. The primary loss function in this stage is imposed on the estimator's output $\hat{\mathbf{Z}}$, i.e., the fusion module, and is defined as the MSE between $\hat{\mathbf{Z}}$ and the ground-truth channel tensor: 
\CheckRmv{
  \begin{equation}
    \mathcal{L}_{\rm main} = \operatorname{MSE}(\hat{\mathbf{Z}}, \mathbf{Z}). \label{eq:loss_fusion}
  \end{equation}
}
To further regularize the pilot processing network, an auxiliary loss function $\mathcal{L}_{\rm aux}$ is introduced to measure the discrepancy between pilot-based estimates $\hat{\mathbf{Z}}_{\rm P}$ and the ground-truth channel tensor, defined as 
\CheckRmv{
  \begin{equation}
    \mathcal{L}_{\rm aux} = \operatorname{MSE}(\hat{\mathbf{Z}}_{\rm P}, \mathbf{Z}).
  \end{equation}
} 
The combined loss, $\mathcal{L}_{\rm tot}=\mathcal{L}_{\rm main} + \mathcal{L}_{\rm aux}$, effectively adapts the PFM to channel estimation and provides an initial optimization of the pilot processing branch.

In the second phase, the PFM backbone is frozen to preserve the learned predictive priors, while the remaining modules are further refined. With the number of trainable parameters reduced, a larger learning rate $l_2$ and a smaller weight decay factor $\lambda_2$ are employed to enable a thorough finetuning. This stage relies solely on the fusion loss \eqref{eq:loss_fusion} to optimize the lightweight components, thereby enhancing estimation performance with improved training efficiency.}

\section{{Numerical Results and Discussions}}  \label{sec:results}  

\subsection{Experimental Details} \label{sec:simu_setup}

\CheckRmv{
  \begin{table}[t]
    \centering
    \caption{Parameter Configurations of the Channel Dataset}
    \begin{threeparttable}
      \begin{tabular}{l|l}
      \toprule
      \textbf{Parameters} & \textbf{Possible Values} \\
      \midrule
      Antenna configurations & BS: 4 antennas; UE: single antenna\\
      Delay profile & \{TDL-A30ns, TDL-B100ns, TDL-C300ns\} \\
      MIMO spatial correlation & \{Low, Medium, Medium-A, High\}\\
      Carrier frequency & 3.5 GHz\\
      Subcarrier spacings &  \{15, 30\} kHz \\
      Number of RBs & 8 (96 subcarriers) \\
      User speeds & \{30, 90, 300\} km/h \\
      SNRs & \{5, 10, 15, 20, 25\} dB \\
      Pilot pattern & \{2P, 4P\} \\
      \bottomrule
      \end{tabular}
    \label{tab:simu_para}

    \end{threeparttable}
  \end{table}
}

\subsubsection{Simulation Settings}
The system configurations and simulation settings for channel dataset generation are summarized in \tabref{tab:simu_para}. The simulation parameters are randomly sampled from the specified ranges in the table to ensure diverse channel conditions. Each slot consists of $T = 14$ OFDM symbols. A total of 8 RBs are allocated, resulting in $K = 96$ subcarriers per slot. To emulate realistic propagation environments, the tapped delay line (TDL) channel model compliant with the 3GPP specifications \cite{3gpp38901} is employed for dataset generation {using the MATLAB 5G toolbox. Specifically, the TDL-A, B, and C profiles represent pure non-line-of-sight (NLOS) conditions with Rayleigh fading taps, while the TDL-D profile, which is utilized later for the out-of-distribution zero-shot generalization evaluation, features a line-of-sight (LOS) Rician fading first tap with a fixed K-factor of 13.3 dB.} Sparse pilot configurations employing FDM are adopted in the simulations, as depicted in the top-left corner of \figref{fig:overall}. The ``2P'' and ``4P'' patterns denote two different allocations of pilot OFDM symbols under $\nt=4$, with $T_{\rm p}=2$ and $T_{\rm p}=4$, respectively, used for different mobility scenarios. These pilot arrangements follow the demodulation RS (DMRS) configuration type 2 defined in 5G NR \cite{dahlman20205g}. 

In total, 768,000 channel trajectories are generated, with each trajectory comprising 10 consecutive slots. {To capture authentic temporal correlations, the channel within each trajectory evolves continuously over time according to the standard 3GPP time-evolution procedure using a Jakes-type Doppler spectrum determined by the designated user speeds and carrier frequency, rather than being independently generated per slot.}
 {This results in} a dataset of 7,680,000 samples, each comprising a triplet of previous and current channel realizations. The dataset is partitioned into training, validation, and test sets using an 8:1:1 split ratio. {It is worth noting that the dataset includes the initial slot configurations where no historical information is available, i.e., $\mathbf{Z}_{\rm his}=\emptyset$. These cold-start samples ensure that the pilot processing network is adequately trained to independently recover the channel when predictive priors are absent.}

\subsubsection{NN Hyperparameters and Training Details} \label{sec:hyperparameter}
For the PFM, we adopt $L_{\rm PFM}=2$ for the backbone to maintain a moderate model size and truncate the first two layers of the pre-trained weights as the initialization of the backbone, a strategy empirically verified to be effective. The latent dimension of the backbone and the number of attention heads are set as $d=1{\rm ,}280$ and $N_{\rm H}=16$, respectively. Following \cite{das2024decoder}, the patch length is set to $L_{\rm pat}=32$ to maximize the prediction performance. Consequently, each input sequence, representing the estimated CSI within one slot, is zero-padded to a length of 32. 
For the ViT-based pilot processing network, the time and frequency domain average lengths are configured as $R_{\rm T}=2$ and $R_{\rm F}=6$, respectively. The number of transformer layers, attention heads, and embedding dimensions are selected as $L_{\rm ViT}=10$, $N_{\rm H}^{\prime}=4$, $d_{\rm m}=128$, and $d_{\rm m}^{\prime}=256$, respectively. For {model training}, the learning rate and weight decay factor of the Adam optimizer are set to $l_1=1\times 10^{-5}$, $\lambda_1=1\times 10^{-2}$, $l_2 =1\times 10^{-4}$, and $\lambda_2 =1\times 10^{-4}$. {The PFM adaptation step and the subsequent two training phases are each conducted for 100 epochs with a batch size of 3,072.} 

\CheckRmv{
  \begin{figure}[t]
    \centering
    \includegraphics[width=3.0in]{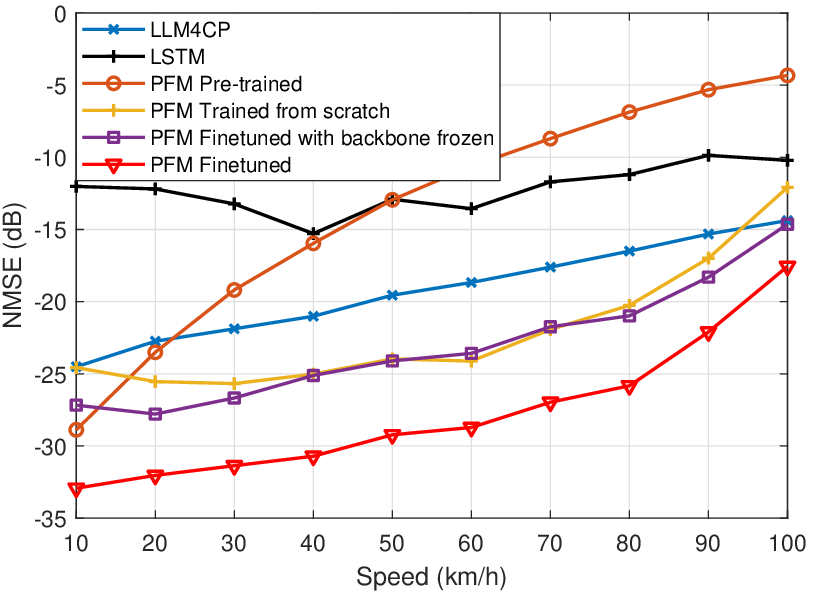}
    \caption{{Channel prediction NMSE performance comparison across varying user speeds. The training and evaluation datasets are adopted from \cite{liu2024llm4cp}, with historical and future sequence lengths set to 16 and 4, respectively.}} 
    \label{fig:pfm_pred}
  \end{figure}
}

\CheckRmv{
  \begin{figure*}[t]
    \centering
      \subfigure[{30 km/h}]{
        \includegraphics[width=2.2in]{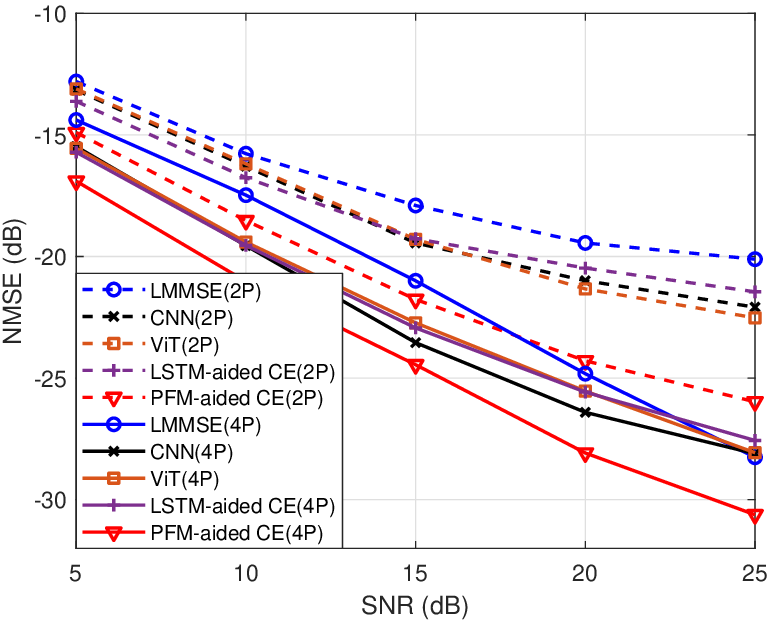}
        \label{fig:nmse_speed1}
      }
      \subfigure[{90 km/h}]{
        \includegraphics[width=2.2in]{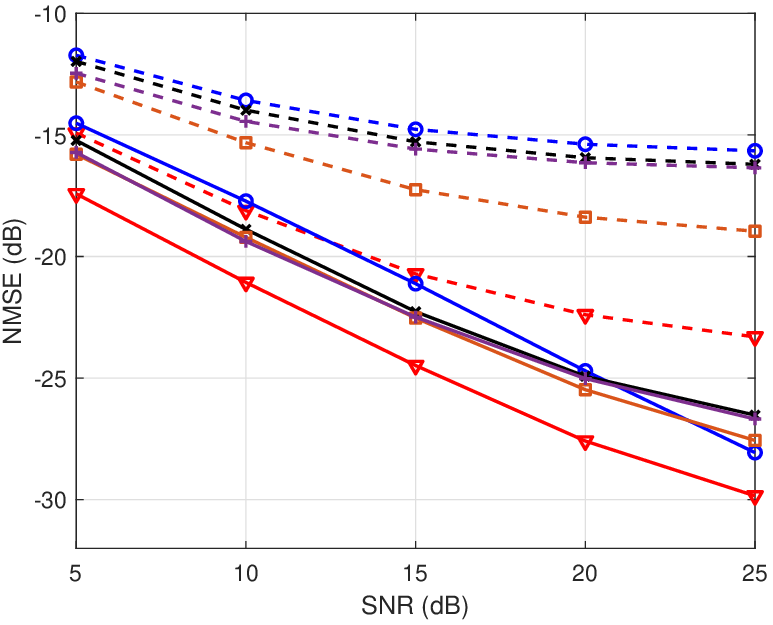}
        \label{fig:nmse_speed2}
      }
      \subfigure[{300 km/h}]{
        \includegraphics[width=2.2in]{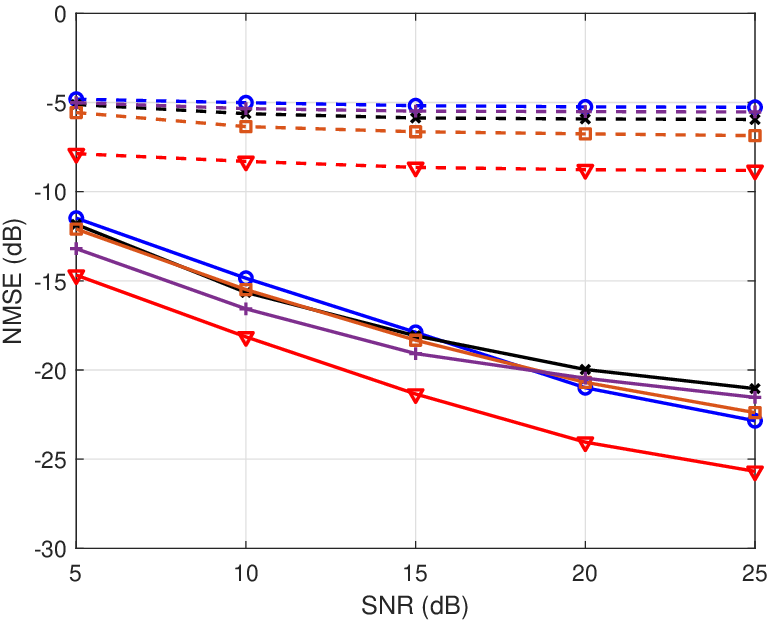}
        \label{fig:nmse_speed3}
      }
    \caption{{NMSE performance with respect to SNRs across varying mobility conditions.}}
    \label{fig:nmse_speed}
  \end{figure*}
}

\subsubsection{Baselines}
The following baselines are compared: 
\begin{itemize}
  \item \textbf{LMMSE:} LS channel estimation with frequency-time domain LMMSE interpolation. The frequency-time domain covariances are computed based on the training dataset for various delay profiles, user speeds, and pilot patterns to ensure robust interpolation performance across diverse channel conditions. 
  \item \textbf{CNN:} Residual network-based CNN channel denoiser \cite{li2019deep}, which takes the LMMSE estimate as input. This model has a total of 1.04M parameters. It is experimentally observed that increasing the network depth or parameter count yields negligible performance gains. 
  \item \textbf{ViT:} This baseline corresponds to the ViT-based model from the proposed approach, configured with identical hyperparameters ($L_{\rm ViT}$, $N_{\rm H}^{\prime}$, $d_{\rm m}$ and $d_{\rm m}^{\prime}$). Similar to the CNN {baseline}, empirical results show that expanding the network depth or feature dimension does not lead to additional performance gains.
  {\item \textbf{LSTM-aided channel estimator (LSTM-aided CE):} A lightweight temporal baseline utilizing a long short-term memory (LSTM) model \cite{liu2024llm4cp,jiang2020deep} to replace the PFM for providing predictive priors. It comprises approximately 1.26M parameters and is trained using the identical training strategy as the proposed approach. This baseline is specifically included to evaluate the necessity of employing a large-capacity foundation model over lightweight recurrent networks.} 
\end{itemize}

\subsection{Prediction Performance of PFM}
{To illustrate the superiority of the PFM and the motivation for its adoption, we compare this model with the state-of-the-art channel prediction method, LLM4CP \cite{liu2024llm4cp}, {and the lightweight LSTM baseline \cite{liu2024llm4cp,jiang2020deep},} as shown in \figref{fig:pfm_pred}. For a fair comparison, the channel datasets released in \cite{liu2024llm4cp} are used for training, finetuning, and performance evaluation of the PFM. To explore effective strategies for leveraging the PFM, four training paradigms are examined: 
\begin{itemize}
  \item \textit{Pre-trained:} The PFM is directly applied using the pre-trained weights\footnote{The pre-trained weights of the PFM developed in \cite{das2024decoder} can be found at \url{https://huggingface.co/google/timesfm-1.0-200m-pytorch}.}, which are obtained from a large-scale time-series corpus, without adaptation. 
  \item \textit{Trained from scratch:} The PFM is randomly initialized and trained solely on the channel datasets, serving as a baseline to assess the benefit of pre-training.
  \item \textit{Finetuned with backbone frozen:} The pre-trained PFM backbone is frozen, while only input and output adapters are finetuned, assessing lightweight domain adaptation.  
  \item \textit{Finetuned:} The entire PFM is finetuned end-to-end on the channel datasets, enabling full domain-specific optimization.
\end{itemize}

The comparative results, depicted in Fig.~\ref{fig:pfm_pred}, lead to several important observations. \textit{First}, the finetuned PFM substantially outperforms the pre-trained version, indicating that adaptation using wireless channel data effectively enhances model performance in channel acquisition. \textit{Second}, when comparing models trained from scratch with those finetuned from pre-trained initialization, the results highlight the advantages of leveraging pre-trained weights from the general time-series foundation model \cite{das2024decoder}. \textit{Finally}, although freezing the backbone during finetuning achieves competitive performance, updating the backbone parameters yields additional gains {and makes the model significantly outperform the LLM4CP baseline}, underscoring the importance of jointly refining the entire model to adapt the PFM to wireless channel-specific tasks.
These observations substantiate the training strategy proposed in Section~\ref{sec:train} for effectively adapting the PFM to wireless channel estimation.}
{Furthermore, the lightweight LSTM baseline significantly lags behind most PFM variants, indicating that conventional lightweight recurrent architectures lack the representational capacity required to capture the complex temporal dependencies and non-linear dynamics inherent in wireless channels.}

\subsection{Channel Estimation Performance Analysis} 
{Unless otherwise specified (e.g., the slot-wise analysis in \figref{fig:nmse_stage}), the performance metrics reported in the following evaluations represent the steady-state performance that is averaged over slot indices $i\geq 2$, excluding the cold-start initial slot. This evaluation accurately reflects the steady-state capability and the specific performance gains introduced by the PFM-aided framework.}

\subsubsection{In-Distribution Performance} %
We first evaluate the in-distribution performance of the proposed PFM-aided channel estimation scheme (PFM-aided CE) compared to baselines and explore the model's ability to handle varying configurations, including user speeds and pilot configurations. The channel estimation NMSE performance across different user velocities is presented in \figref{fig:nmse_speed}. As shown in the figure, PFM-aided CE outperforms all baselines by 1--5 dB in NMSE across all velocities and SNRs under both the ``2P'' and ``4P'' pilot configurations. This gain primarily stems from the model's ability to effectively exploit temporal knowledge from previously estimated CSI. Specifically, at moderate speeds of 30 km/h and 90 km/h, the PFM-aided CE with the sparse ``2P'' configuration even achieves {performance comparable to} LMMSE using the denser ``4P'' configuration, especially in the low-SNR regime. 
{It is worth mentioning that the exceptional results for both the ``2P'' and ``4P'' configurations are achieved using the same pre-trained model, validating the architectural scalability of the proposed model and eliminating the overhead of pattern-specific retraining.}
{At high mobility (300 km/h), the temporal correlation of the channel becomes significantly low, leading to severe channel dynamics. As expected, all compared approaches experience performance degradation. However, the proposed estimator still delivers the best performance and achieves even more pronounced gains. These results demonstrate that the PFM can extract complex temporal dependencies even when linear temporal correlation is weak, confirming the ability of the model to handle channel variations with different time granularities.}

{Moreover, regarding the impact of temporal model capacity, the lightweight LSTM-aided CE significantly underperforms the proposed framework, as shown in \figref{fig:nmse_speed}. Crucially, under the sparse ``2P'' configuration, inaccurate LSTM predictions introduce a negative temporal gain, degrading the estimation accuracy even below that of the standalone, non-predictive ViT. This performance penalty highlights that lightweight models fail to provide reliable predictive priors, confirming the essential role of the PFM for high-fidelity channel acquisition.}

\CheckRmv{
  \begin{figure}[t]
    \centering
    \includegraphics[width=3.0in]{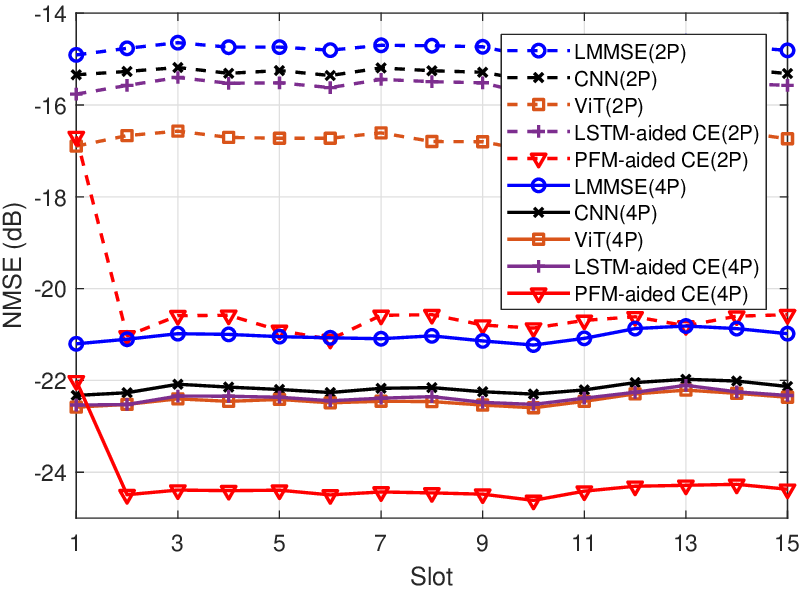}
    \caption{{NMSE performance with regard to the slot index under the speed of 90 km/h and \SNR{=}{15}.}}
    \label{fig:nmse_stage}
  \end{figure}
}

{To evaluate the long-term stability of the recursive inference workflow and address potential concerns regarding exposure bias, i.e., the mismatch between using LMMSE estimates during training and the model's own outputs during inference, we extend the evaluation trajectory to 15 consecutive slots, well beyond the 10-slot trajectory length used during training. The NMSE performance with regard to the slot index is presented in \figref{fig:nmse_stage}, where a moderate speed of 90 km/h and SNR of 15 dB are considered. The results show that the proposed estimator maintains stable performance across multiple slots for slot indices $i\geq 2$, clearly outperforming the first slot, which relies solely on the pilot observations for estimation. This observation highlights the effectiveness of the proposed ``predict-and-refine'' framework.
Particularly, PFM-aided CE with ``2P'' shows virtually the same estimation accuracy as LMMSE with ``4P'', despite using only 50\% of the pilot overhead, demonstrating its significant potential in reducing pilot overhead.

Furthermore, the absence of error divergence over the 15-slot continuous operation validates the framework's resilience against error propagation.
Unlike pure autoregressive generation where errors can compound, instantaneous pilot observations at each slot serve as anchors to systematically calibrate the predictive priors. Additionally, training the PFM on imperfect LMMSE estimates inherently equips the model to handle noisy historical contexts, ensuring stable adaptation to its own outputs during continuous inference.}

\CheckRmv{
  \begin{figure}[t]
    \centering
    \includegraphics[width=3.0in]{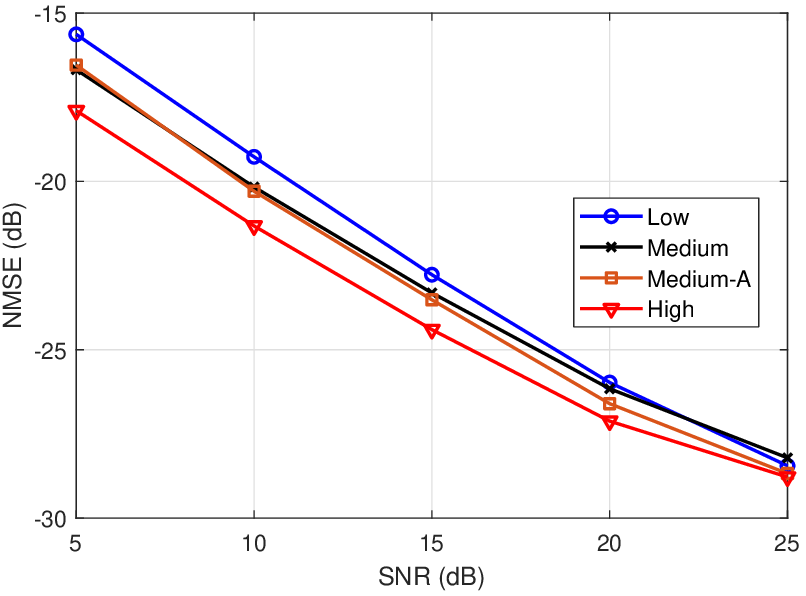}
    \caption{{NMSE performance of the proposed PFM-aided CE across different spatial correlation levels.}}
    \label{fig:nmse_correlation}
  \end{figure}
}

{Next, we investigate the effectiveness of the proposed architecture under varying degrees of spatial correlation. 
It is worth recalling that the PFM branch employs a univariate decomposition strategy, which treats the sequence of each antenna independently. To verify whether the proposed architecture successfully compensates for this lack of spatial awareness during the prediction phase, we evaluate the estimation NMSE across different spatial correlation levels defined in the 3GPP specifications (Low, Medium, Medium-A, and High) \cite{3gpp36101}. 
As illustrated in \figref{fig:nmse_correlation}, the estimation accuracy improves as the spatial correlation increases. This trend demonstrates that rather than discarding spatial mutual information across antennas, the proposed ``predict-and-refine'' framework effectively captures it through the pilot processing ViT. 
For spatially correlated channels, the fusion module successfully exploits this rich spatial context to calibrate the independent temporal predictive priors provided by the PFM, thereby achieving superior overall channel estimation performance.}

\subsubsection{Zero-Shot Generalization Study}  
\CheckRmv{
  \begin{figure}[t]
    \centering
    \includegraphics[width=3.0in]{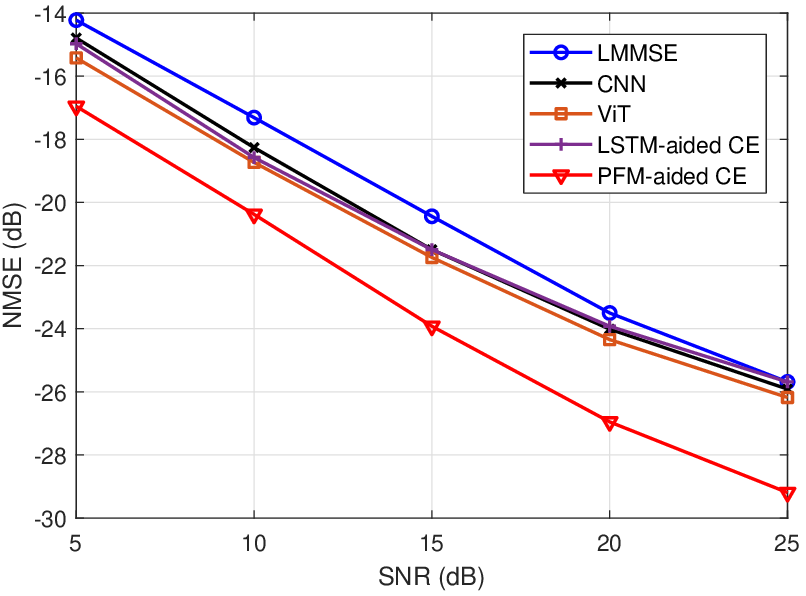}
    \caption{{Zero-shot generalization performance under the unseen user speed of 150 km/h.}}
    \label{fig:unseen_speed}
  \end{figure}
}
To investigate the generalization capability of the proposed estimator, we evaluate performance under unseen scenarios, including {different user velocities, antenna configurations}, and channel environments, without any additional training. \figref{fig:unseen_speed} presents the NMSE results under an unseen mobility condition with a user speed of 150 km/h. In this high-mobility scenario, all compared estimators employ the ``4P'' pilot pattern to track the rapidly time-varying channels. Among the compared approaches, the proposed estimator exhibits the strongest generalization performance. Specifically, it outperforms all compared baselines by over 2 dB in estimation NMSE at \SNR{=}{20}. {This superiority indicates that the model successfully learns generalizable temporal representations of the channel dynamics}, enabling reliable channel acquisition accuracy even when the temporal correlations of the testing channels differ from those observed during training. 

\CheckRmv{
  \begin{figure}[t]
    \centering
    \includegraphics[width=3.0in]{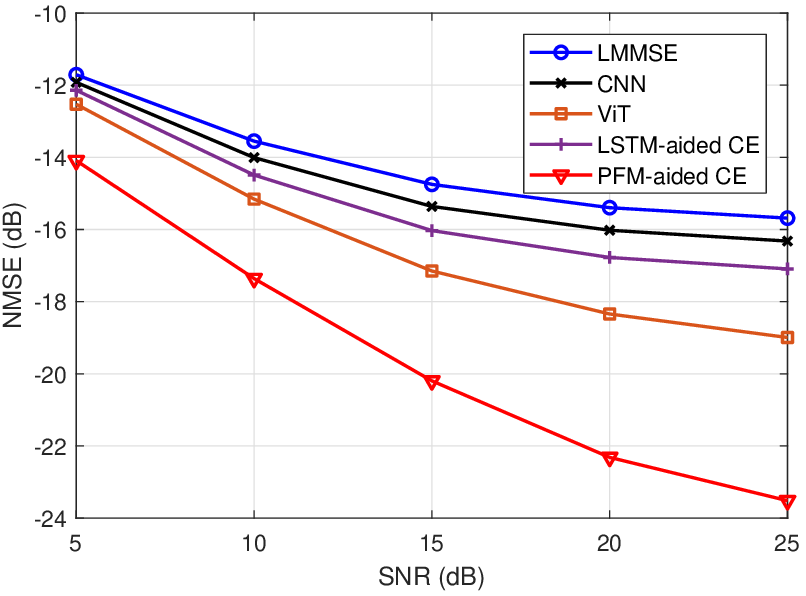}
    \caption{{Zero-shot generalization performance when the BS is equipped with 16  antennas ($\nt=16$).}}
    \label{fig:unseen_ant}
  \end{figure}
}
To investigate the scalability and generalizability of the proposed scheme to varying antenna configurations, \figref{fig:unseen_ant} shows the channel estimation performance when the BS is equipped with 16 transmit antennas, a configuration unseen during training. Owing to the scalable architecture design, featuring univariate decomposition and AdaLN, the model can be directly applied to this unseen setting without any modification. The results also showcase that the model generalizes well, outperforming all baselines by more than 4 dB in NMSE for \SNR{\geq}{20}. Similar observations are obtained when varying other system parameters, such as the number of subcarriers or RBs.
Notably, as the antenna count increases, the achieved pilot overhead reduction  becomes more pronounced under this scenario, which is crucial for enhancing transmission efficiency. This advantage is particularly valuable for future wireless networks, where the dimensionality of antenna arrays continues to increase.

\CheckRmv{
  \begin{figure}[t]
    \centering
    \includegraphics[width=3.0in]{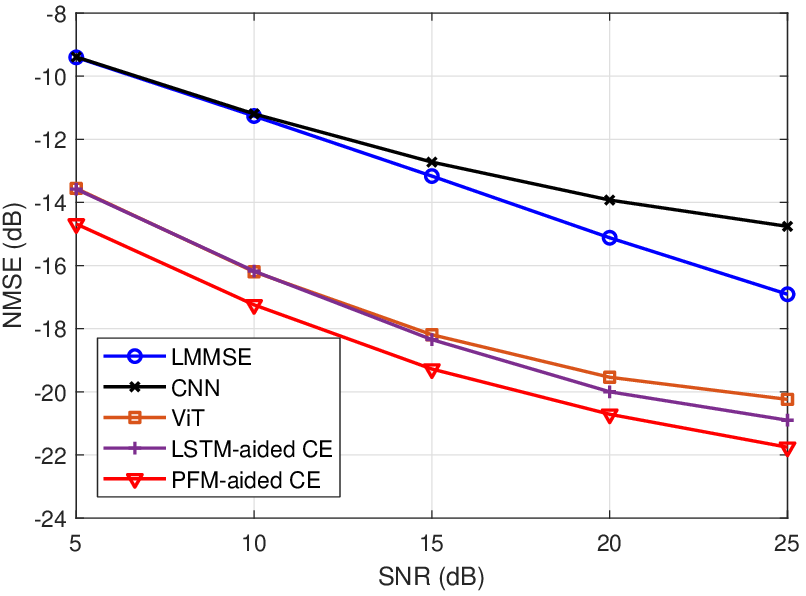}
    \caption{{Zero-shot generalization performance under the unseen TDL-D channel profile.}}
    \label{fig:unseen_scenario}
  \end{figure}
}
We further investigate the NMSE performance under an unseen environment using the LOS TDL-D profile, as shown in \figref{fig:unseen_scenario}. This setting introduces a substantial distributional shift in channel correlation and delay structure, as the model was trained exclusively on NLOS scenarios (TDL-A/B/C). {Despite this drastic mismatch}, the proposed PFM-aided CE maintains competitive performance, remarkably surpassing {all the baselines}. These results underscore the model's zero-shot generalization ability, enabling adaptation to structurally different channel environments without finetuning. {This capability stems from} the time-series forecasting knowledge inherited from the pre-trained PFM. This large-capacity model has been pre-trained across a large-scale time-series corpus, thereby generalizing effectively to out-of-distribution wireless channel sequences, an ability inaccessible to conventional statistical or small-scale AI models.

\subsection{{End-to-End Bit Error Rate (BER) Performance}} 

\CheckRmv{
  \begin{figure}[t]
    \centering
      \subfigure[{90 km/h, 2P}]{
        \includegraphics[width=3.0in]{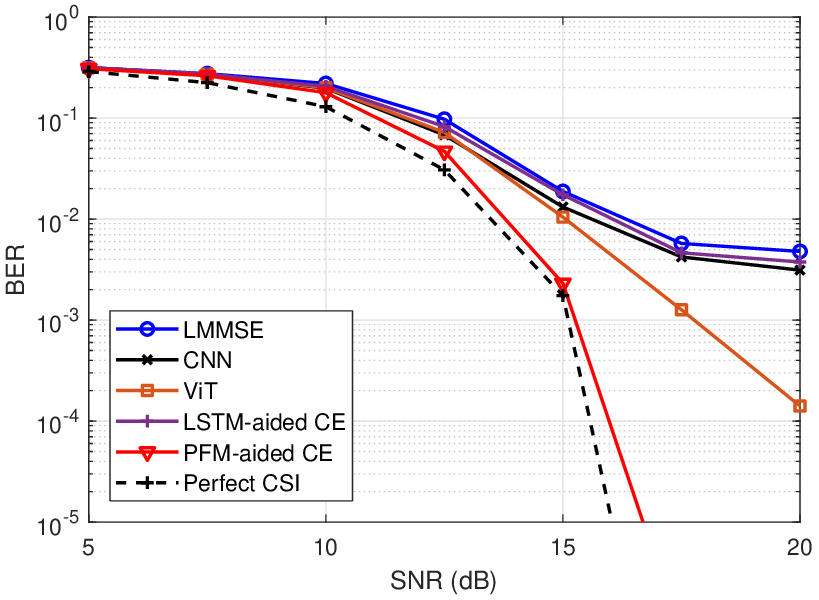}
        \label{fig:ber_speed1}
      }
      \subfigure[{300 km/h, 4P}]{
        \includegraphics[width=3.0in]{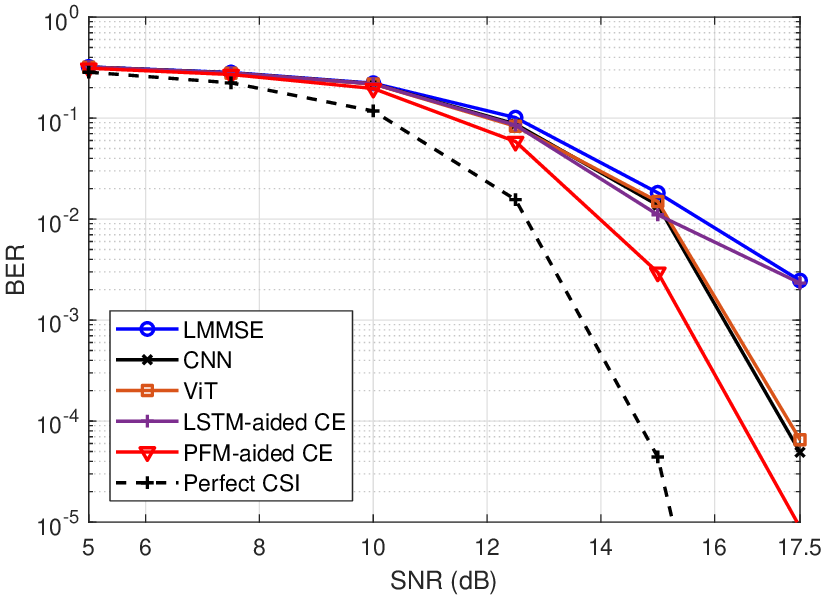}
        \label{fig:ber_speed2}
      }
    \caption{{End-to-end BER performance comparison of the communication link employing the proposed channel estimator and various baseline methods.}}
    \label{fig:ber}
  \end{figure}
}

To assess the practical viability of the proposed channel estimator in a communication link, we evaluate its impact on end-to-end BER. We consider a multiple-input single-output OFDM system, and the channel settings follow those presented in \tabref{tab:simu_para}. The 5G NR low-density parity-check (LDPC) code with a rate of 466/1024 \cite{3gpp38214} and a codeword size of 7296 bits 
is employed to encode the message bits for all data REs. The encoded bits are then modulated using 64-ary quadrature amplitude modulation (64-QAM) and mapped onto the data REs over the resource grid. 
The data stream is precoded for transmission across $\nt=4$ antennas. 
{The transmitter is assumed to perform precoding based on the estimated CSI perfectly fed back from the user. The digital precoder on the $(k,t)$-th RE, $\mathbf{v}_{k,t} \in \mathbb{C}^{4\times 1}$, is obtained based on match filtering \cite{tse2005fundamentals} using the estimated full CSI $\hat{\mathbf{h}}_{k,t} \in \mathbb{C}^{4\times 1}$ on the $(k,t)$-th RE:
\CheckRmv{
    \begin{equation}
    \mathbf{v}_{k,t} = \frac{\hat{\mathbf{h}}_{k,t}}{\|\hat{\mathbf{h}}_{k,t}\|_2}.
    \label{eq:mf}
  \end{equation}
}
Hence, the received signal at the $(k,t)$-th RE is given by
\CheckRmv{
  \begin{equation}
    y_{k,t} = \mathbf{h}_{k,t}^{\rm H} \mathbf{v}_{k,t}x_{k,t} + n_{k,t},
  \end{equation}
}
where $\mathbf{h}_{k,t} \in \mathbb{C}^{4\times 1}$ represents the true CSI, and $x_{k,t}$ and $n_{k,t}$ denote the transmitted symbol and the AWGN with variance $\sigma^2$ at the $(k, t)$-th RE, respectively.}

{At the receiver side, the received signals are processed using an LMMSE equalizer. To isolate the evaluation to the accuracy of the proposed estimator, the equivalent channel required for equalization is directly reconstructed by the UE using its previously estimated CSI, i.e., $\hat{\mathbf{h}}_{k,t}^{\rm H} \mathbf{v}_{k,t} = \|\hat{\mathbf{h}}_{k,t}\|_2$.} The equalized signals are then soft-demapped and delivered to the LDPC decoder to recover the transmitted message bits. The BER is evaluated after $10^4$ codewords are transmitted. 

{The BER results obtained using different channel estimators under various mobility conditions are presented in \figref{fig:ber}. For reference, the BER performance with perfect CSI at {both the transmitter and the receiver} is included as an upper bound. \figref{fig:ber_speed1} presents the results for a user speed of 90 km/h with the ``2P'' pilot pattern. As shown in the figure, the proposed estimator enables a substantial BER reduction in the high-SNR regime, {surpassing all the practical baselines by approximately 4 dB} and closely approaching the upper bound. These results highlight the critical role of the accurate channel acquisition achieved by the proposed PFM-aided CE in enhancing the overall system performance with only limited pilot overhead. \figref{fig:ber_speed2} further evaluates the estimators at a higher user speed of 300 km/h using the ``4P'' pilot pattern. Although the performance gap between the proposed scheme and the upper bound becomes larger due to increased channel dynamics, the proposed method continues to outperform all baselines by a significant margin, demonstrating strong robustness under high-mobility conditions.}

\subsection{Ablation Study}
\CheckRmv{
  \begin{figure}[t]
    \centering
    \includegraphics[width=3.0in]{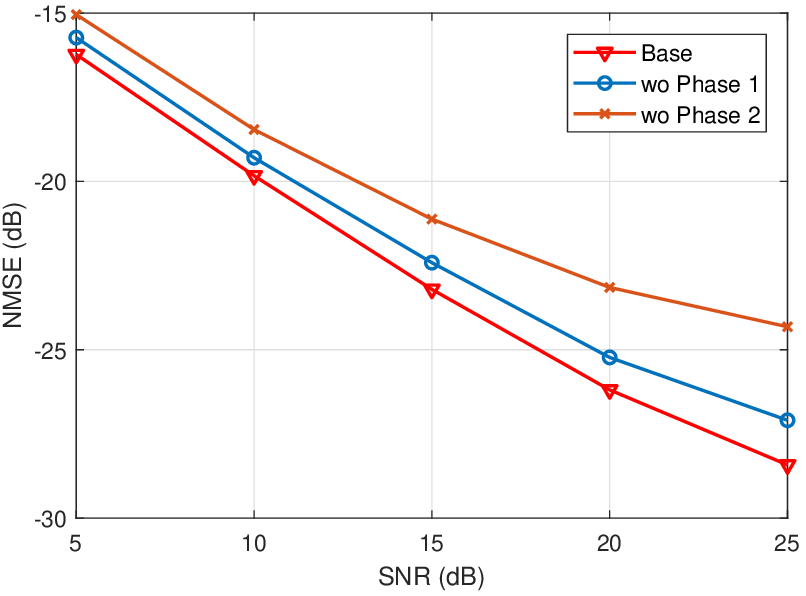}
    \caption{{Ablation study on the proposed two-phase training strategy.}}
    \label{fig:ablation_wophase}
  \end{figure}
}
In this subsection, we conduct ablation experiments to evaluate the {contributions} of several critical components and design strategies within the proposed channel estimator.\footnote{The ablation of modules in the PFM for channel prediction, such as positional encoding and patching, has been reported in \cite{sheng2025wireless}. Based on those results, we omit repetitive experiments for channel estimation in this study.} First, we examine the proposed two-phase training strategy by selectively removing each phase, as demonstrated in \figref{fig:ablation_wophase}. It can be observed that omitting either phase (wo Phase 1 or wo Phase 2) leads to a notable performance drop. Moreover, a comparison of the two phase-removal baselines indicates that Phase 2 plays a more critical role in performance enhancement, as this stage enables sophisticated adaptation of the model to wireless channel estimation tasks.  

\CheckRmv{
  \begin{figure}[t]
    \centering
    \includegraphics[width=3.0in]{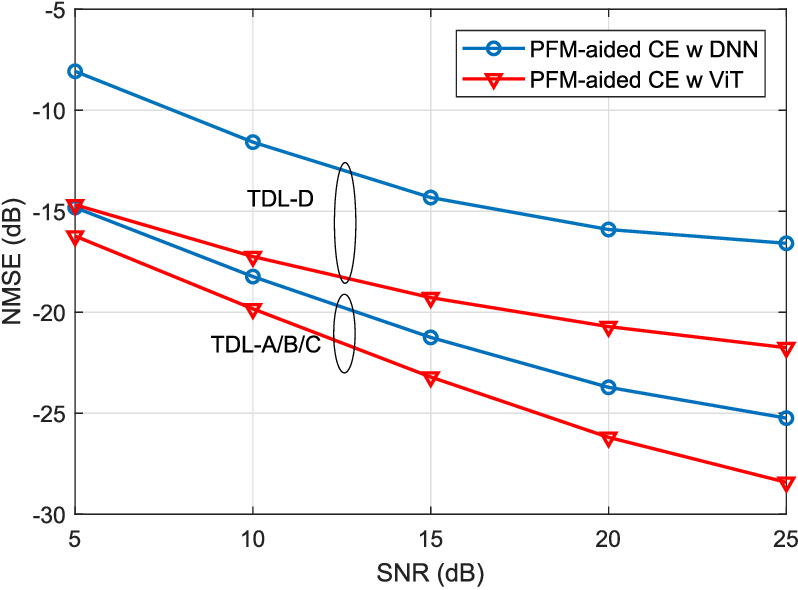}
    \caption{{Ablation comparison between using the proposed ViT structure and a DNN-based alternative for pilot information processing.}}
    \label{fig:ablation_dnn_vs_vit}
  \end{figure}
}
{To} validate the effectiveness of the proposed pilot processing ViT, we conduct an ablation study by replacing this module with a two-layer FC deep NN (DNN) that maps the initial estimate $\hat{\mathbf{H}}_{\rm P}$ to a hidden state of dimension 1280 for fusion. \figref{fig:ablation_dnn_vs_vit} shows the compared results, where the PFM-aided CE using DNN for pilot processing exhibits substantially inferior performance, particularly in terms of generalization to the out-of-distribution TDL-D channel profile. This comparison highlights that the proposed ViT for capturing spatial, frequency, and temporal CSI correlations is essential for the high-fidelity channel acquisition achieved by our framework.   

\CheckRmv{
  \begin{figure}[t]
    \centering
    \includegraphics[width=3.0in]{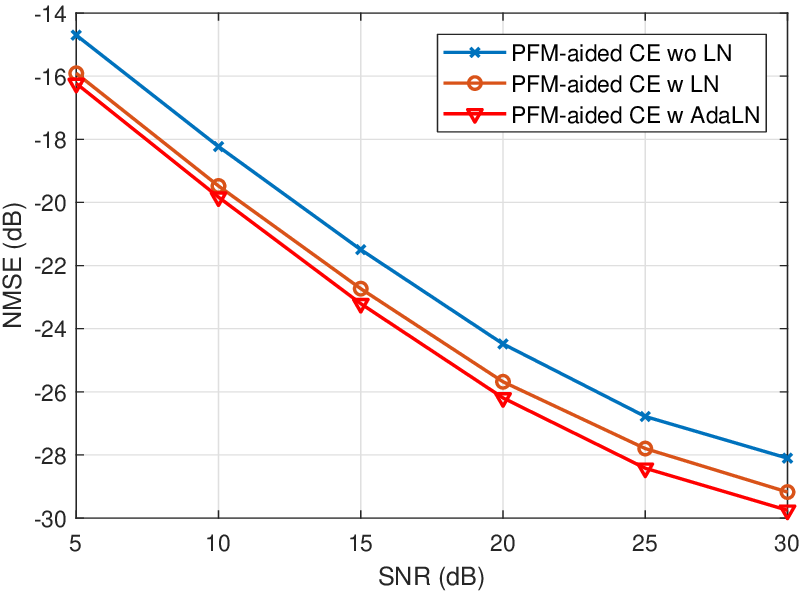}
    \caption{{Ablation study on layer normalization within the enhanced FFN of the pilot processing ViT.}}
    \label{fig:ablation_ln}
  \end{figure}
}
Finally, we investigate the impact of different layer normalization strategies within the pilot processing ViT's enhanced FFN. As shown in \figref{fig:ablation_ln}, the absence of layer normalization (``wo LN'') results in a significant performance drop. Moreover, using depth-wise AdaLN yields comparable and even superior performance compared to conventional dimension-wise layer normalization with fixed scale and shift parameters (``w LN''). In addition to its performance advantages, AdaLN provides scalability and adaptability to different system dimensions that are not attainable by the ``w LN'' baseline.

\vspace{-0.1cm}
\subsection{Complexity Analysis} 

\CheckRmv{  
  \begin{table}[t]
    \centering
    \caption{Complexity Comparison in terms of Inference Latency and Network Parameters}
     \begin{threeparttable}
      \begin{tabular}{lcc}
      \toprule
      \textbf{Methods} & \textbf{Inference Latency [ms]} & \textbf{Parameters [M]} \\
      \midrule
      LMMSE & 0.43  &  / \\
      CNN & 0.52  & 1.04 \\
      ViT & 0.83  & 3.13 \\
      {LSTM-aided CE} & {1.60}  & {4.39}\\
      PFM-aided CE & 2.81  & 24.00   \\
      \bottomrule
      \end{tabular}
    \label{tab:complexity}
    \end{threeparttable}
  \end{table}
}
In this subsection, we analyze the complexity of the proposed estimator, including the inference latency per slot and the parameter count, to {assess its practicality for real-world deployment}. The complexity results, as shown in \tabref{tab:complexity}, {were obtained using} a machine with two NVIDIA GeForce RTX4090 GPUs and an Intel Core i9-14900K CPU. The inference latency is measured under the settings of \tabref{tab:simu_para} for each slot and averaged over 100 channel uses. The table demonstrates that the latency of the proposed approach remains at an acceptable level, well within 3 ms. This result is close to the latency requirement specified in the 3GPP standard \cite{3gpp38802}, and it can be further optimized by designing application-specific computing architectures for foundation models. 

{Regarding} model complexity, the proposed estimator contains a substantially larger number of parameters compared to baseline models, primarily due to the high-dimensional latent space of the PFM's transformer backbone. This increased parameter count is fundamental for the model's capacity and powerful generalization, as demonstrated in the simulation results. That being said, the total parameter count remains significantly lower than that of the original PFM released in \cite{das2024decoder} (200M parameters), making it feasible for deployment on typical communication equipment.

\subsection{{Scalability and Implementation Feasibility}}

\CheckRmv{
  \begin{table}[t]
    \centering
    \caption{{Complexity Extension of the Proposed PFM-aided CE for Scaled Antenna Scenarios}}
    \begin{threeparttable}
    \label{tab:scalability}
    {\begin{tabular}{llcc}
        \toprule
        \textbf{Scenarios} & \textbf{Number} & \textbf{Latency [ms]} & \textbf{Parameters [M]}  \\
        \midrule
        \multirow{4}{*}{\shortstack[l]{\textbf{Varying Effective} \\ \textbf{Tx Ports}}} 
        & 4 & 2.81 & 24.00 \\
        & 8 & 4.61 & 24.00 \\
        & 16 & 9.29 & 24.00 \\
        & 32 & 17.86 & 24.00 \\
        \midrule
        \multirow{3}{*}{\shortstack[l]{\textbf{Varying User} \\ \textbf{Rx Antennas}}} 
        & 1 & 2.81 & 24.00 \\
        & 2 & 3.17 & 24.00 \\
        & 4 & 5.75 & 24.00 \\
        \bottomrule
    \end{tabular}}
    \end{threeparttable}
\end{table}  
}

{To comprehensively assess the practical viability of the proposed framework, we extend our analysis to scaled antenna and multi-user configurations and discuss its hardware implementation feasibility.

\subsubsection{Scalability in Large-Scale MIMO and Multi-User Systems}
First, in practical large-scale MIMO downlinks, the UE estimates the effective precoded antenna ports rather than hundreds of physical antennas. When scaling the number of effective ports $\nt$, the proposed model maintains a strictly constant parameter count of 24.00M, thereby avoiding the need for retraining. This is achieved through the univariate decomposition in the PFM and the AdaLN in the ViT. Computationally, the PFM scales linearly, $\mathcal{O}(\nt)$, as it processes independent sequences, while the ViT's spatial self-attention scales quadratically as $\mathcal{O}(\nt ^2)$. Because the practical number of effective ports is bounded and the linear-complexity PFM dominates the computation load, the overall latency experiences highly manageable growth without severe quadratic explosion.
    
Furthermore, for the downlink scenario, channel estimation is executed locally at the UE. Because this computation is fully decentralized, serving numerous UEs within a BS's coverage area does not increase the computational burden on any individual UE. When scaling to multi-antenna UEs, the channel tensor dimensions expand. To maintain scalability, the proposed framework can process multiple receive antennas as parallel independent sequences rather than performing joint cross-antenna attention. This design choice circumvents the quadratic complexity explosion associated with self-attention mechanisms and also maintains the model parameter count at 24.00M. As summarized in \tabref{tab:scalability}, whether scaling the effective transmit ports or user receive antennas, the zero-parameter-increase property demonstrates exceptional architectural scalability suitable for practical deployments, with inference latency scaling linearly or sub-linearly due to batch parallelization.
Finally, as noted in the system model, while formulated within a downlink context to simplify the mathematical presentation, our framework can be seamlessly deployed at the BS for uplink multi-user estimation. For such deployments, the framework can process multiple users via parallel execution, similar to the multi-antenna UE case.

\subsubsection{Implementation Feasibility} 
Deploying the proposed model on a power-constrained UE necessitates specific hardware considerations. 
First, it is worth noting that the 24M-parameter footprint is relatively moderate, especially when compared to the recent trend of deploying lightweight LLMs with sub-billion to billion parameters on mobile edge devices \cite{liu2024mobile}.
This footprint can be substantially compressed using established techniques such as INT8 quantization and weight pruning, allowing the model to efficiently reside within mobile memory hierarchies while drastically cutting power consumption.
Furthermore, to meet the stringent real-time latency of the PHY layer, practical implementations can leverage the powerful neural processing units (NPUs) \cite{raha2025llm}  increasingly integrated into modern mobile system-on-chips, which are specifically optimized for highly parallel workloads as envisioned in 3GPP Release 18/19 AI-native PHY layer designs \cite{lin2025ai}. Finally, as clarified in the previous subsection, the proposed estimator is seamlessly applicable to BS-side uplink channel estimation, where abundant computing resources are available, making the deployment of such foundation models highly feasible.}

\section{Conclusion} \label{sec:conclusion}
In this paper, we proposed a PFM-aided wireless channel acquisition framework that unifies predictive inference and pilot-based estimation into a cohesive and scalable architecture. By leveraging the strong prior knowledge captured by large-scale foundation models and incorporating pilot observations, the proposed method enables highly generalizable, scalable, and context-aware channel estimation. Moreover, a tailored pilot processing network architecture was designed to effectively learn spatial-temporal-frequency correlations in CSI, while a predictive-pilot fusion mechanism was introduced to achieve robust reconstruction even under sparse pilot conditions. Extensive experiments validated that the proposed PFM-aided estimator achieves substantial performance gains over both traditional and learning-based baselines in estimation accuracy, robustness, and adaptability across diverse channel scenarios. These results demonstrate the strong potential of PFMs to serve as a cornerstone for intelligent, adaptive, and scalable channel acquisition in future wireless communication systems.

{\appendices

\section{Causal Transformer Details}  \label{app:causal}
The causal transformer comprises stacked layers, each containing a multi-head causal self-attention module and an FFN. The multi-head attention module contains $N_{\rm H}$ scaled-dot product attention heads, of which the input is composed of the query ($\mathbf{Q}$), key ($\mathbf{K}$), and value ($\mathbf{V}$) matrices, denoted as $\mathbf{Q}_{h}=\mathbf{E}\mathbf{W}^{{\rm Q}}_{h}$, $\mathbf{K}_{h}=\mathbf{E}\mathbf{W}^{{\rm K}}_{h}$, $\mathbf{V}_{h}=\mathbf{E}\mathbf{W}^{{\rm V}}_{h}$, respectively, where $h\in \{1,\ldots, N_{\rm H}\}$. $\mathbf{W}^{{\rm Q}}_h,\mathbf{W}^{{\rm K}}_h\in \mathbb{R}^{d \times d_k}$ and $\mathbf{W}^{{\rm V}}_h\in \mathbb{R}^{d \times d_v}$ denote the learnable projection matrices, where $d_k=d_v=d / N_{\rm H}$. After the input transformation, the \textit{causal} attention operation for the $h$-th head is expressed as
\CheckRmv{
  \begin{equation}
     \text{head}_{h} = \operatorname{Softmax}\left(\frac{\mathbf{Q}_{h}\left(\mathbf{K}_{h}\right)^{\top} + \mathbf{M}}{\sqrt{d_k}} \right)\mathbf{V}_{h} \in \mathbb{R}^{N_{\rm pat} \times d_v}, \label{eq:head}
  \end{equation}
}
where $\mathbf{M} \in \mathbb{R}^{N_{\rm pat} \times N_{\rm pat}}$ denotes the causal masking matrix, whose $(m,j)$-th element is given by 
\CheckRmv{
  \begin{equation}
    \mathbf{M}_{(m, j)} = \left\{\begin{array}{ll}
      0, & \text { if } m\geq j, \\
      -\infty, & \text { otherwise},
    \end{array}\right.
  \end{equation}
}
{ensuring that position} $m$ attends to only its current and preceding positions ($j\leq m$), thereby preventing information leakage from future tokens.

Furthermore, the outputs of all attention heads are concatenated and linearly projected as
\CheckRmv{
  \begin{equation}
    \mathbf{O} =
    \left[\text{head}_1, \ldots, \text{head}_{N_{\rm H}}\right]\mathbf{W}^{{\rm O}},
    \label{eq:multi_head}
  \end{equation}
}
where $\mathbf{W}^{{\rm O}} \in \mathbb{R}^{N_{\rm H}d_v\times d}$ aligns the aggregated representation with the latent dimension of the transformer. {The resulting multi-head attention output $\mathbf{O} \in \mathbb{R}^{N_{\rm pat} \times d}$ is then} processed by the FFN to derive the layer output.

\section{Pilot Processing Network Details} \label{app:pilot}
\subsection{Input Feature Preparation}
The despreading operation derives $\nt$ feature maps of size $K \times L_{\rm B}$, denoted as $\hat{\mathbf{H}}_{\rm P}^{\prime}$, where $L_{\rm B} = T / R_{\rm T}$. By concatenating the real and imaginary parts along the antenna dimension, $\hat{\mathbf{H}}_{\rm P}$ and $\hat{\mathbf{H}}_{\rm P}^{\prime}$ are derived as $2\nt$ feature maps of size $K\times T$ and $K \times L_{\rm B}$. These feature maps are concatenated into a feature tensor $\mathbf{A} \in \mathbb{R}^{2\nt \times K \times L_{\rm f}}$, where $L_{\rm f} = T + L_{\rm B}$.

Following the procedure outlined in \cite{dosovitskiy2020image}, the input feature first passes through a patch embedding module prior to the transformer encoder.  
To capture spatial dependencies among antennas, the spatial-time domain channel on each subcarrier, denoted as $\mathbf{A}_k \in \mathbb{R}^{2\nt \times L_{\rm f}}$ ($k \in \{1,\ldots, K\}$), is processed separately to simplify representation learning. Subsequently, each matrix $\mathbf{A}_k$ is partitioned into a sequence of $2\nt$ patches, where each patch corresponds to a feature vector of length $L_{\rm f}$. This patching strategy encourages the model to focus on learning channel spatial correlations across antenna elements. 
The resulting patches are then mapped into a latent space of dimension $d_{\rm m}$ via an FC layer, yielding the patch embeddings $\mathbf{F}_k \in \mathbb{R}^{2\nt\times d_{\rm m}}$ for each subcarrier. 

\subsection{Enhanced FFN Operations}

The enhanced FFN integrates a DWConv operation to capture local frequency correlations.  Denote $H\triangleq K$,  $W\triangleq 2\nt$, and we have $Q= H\cdot W$. Let $\mathbf{R}_{\rm e} \in \mathbb{R}^{Q\times d_{\rm m}}$ denote the input tokens from the attention module. The operations of the enhanced FFN can be summarized as
\CheckRmv{
  \begin{subequations}  
    \begin{align}
    \mathbf{R}_{l_1}&=\operatorname{GELU}\left(\operatorname{AdaLN}\left(\operatorname{Linear1}(\mathbf{R}_{\rm e})\right)\right), \\
    \mathbf{R}_{l_1}&\in\mathbb{R}^{Q\times d_{\rm m}^{\prime}}\to\mathbf{R}_{l_1}^{\prime}\in\mathbb{R}^{H\times W\times d_{\rm m}^{\prime}}, \label{eq:restore}\\
    \mathbf{R}_{\rm d}^{\prime}&=\operatorname{GELU}\left(\operatorname{AdaLN}\left(\operatorname{DWConv}(\mathbf{R}_{l_1}^{\prime})\right)\right), \\
    \mathbf{R}_{\rm d}^{\prime}&\in\mathbb{R}^{H\times W\times d_{\rm m}^{\prime}} \to\mathbf{R}_{\rm d}\in \mathbb{R}^{Q\times d_{\rm m}^{\prime}}, \label{eq:flatten} \\
    \mathbf{R}_{l_2}&=\operatorname{AdaLN}\left(\operatorname{Linear2}\left(\mathbf{R}_{\rm d}\right)\right)+\mathbf{R}_{\rm e}, 
    \end{align}
  \end{subequations}
}
where the {symbol} $\to$ in \eqref{eq:restore} and \eqref{eq:flatten} denotes reshaping operations between the flattened sequence ($Q$) and the 2D spatial-frequency grid ($H \times W$).}



\ifCLASSOPTIONcaptionsoff
  \newpage
\fi




\end{document}